\documentclass[fleqn,usenatbib]{mnras}

\usepackage[T1]{fontenc}
\usepackage{ae,aecompl}
\usepackage{hyperref}

\newcommand{\ebv}{\textit{E(B-V)\/}}

\newcommand{\jb}{\textit{J\/}}
\newcommand{\h}{\textit{H\/}}
\newcommand{\ks}{\textit{K\/}$_\mathrm{s}$}

\newcommand{\gaia}{\textit{Gaia\/}}
\newcommand{\g}{\textit{G\/}}
\newcommand{\gbp}{\textit{G\/}$_\mathrm{BP}$\/}
\newcommand{\grp}{\textit{G\/}$_\mathrm{RP}$\/}

\newcommand{\spitzer}{\textit{Spitzer\/}}

\newcommand{\tmass}{\textit{2MASS\/}}
\newcommand{\wise}{\textit{WISE\/}}

\newcommand{\planck}{\textit{Planck\/}}
\newcommand{\lsr}{\textit{LSR\/}}
\newcommand{\msun}{\,M$_{\sun}$}

\newcommand{\av}{\textit{A\/}$_\mathrm{V}$}

\newcommand{\mincls}{{$\mathtt{minimum\_cluster\_size}$}}
\newcommand{\mins}{{$\mathtt{minimum\_samples}$}}
\newcommand{\csm}{{$\mathtt{cluster\_selection\_method}$}}
\newcommand{\vllsr}{$v_{\mathrm{l,LSR}}$}
\newcommand{\vblsr}{$v_{\mathrm{b,LSR}}$}
\newcommand{\kms}{km\,s$^{-1}$}
\newcommand{\masy}{mas\,y$^{-1}$}

\usepackage{url}

\usepackage{graphicx}	
\usepackage{amsmath}	
\usepackage{amssymb}	
\usepackage{enumitem}
\usepackage{pdflscape}
\usepackage{lipsum}  
\usepackage{appendix}

\title[Cepheus~OB2]{The \textit{Gaia} view of the Cepheus~OB2 association}

\author[Szil\'agyi  et al.]{M\'at\'e Szil\'agyi\textsuperscript{1,2,3}\thanks{E-mail: szilagyi.mate@csfk.org}, M\'aria Kun\textsuperscript{1,2},
P\'eter \'Abrah\'am\textsuperscript{1,2,3}, G\'abor Marton\textsuperscript{1,2,3} \\
$^{1}$Konkoly Thege Mikl\'os Astronomical Institute, Research Centre for Astronomy and Earth Sciences, E\"otv\"os Lor\'and Research Network,\\
H-1121 Budapest, Konkoly-Thege Mikl\'os \'ut 15--17, Hungary\\
${^2}$CSFK,  MTA Centre of Excellence, H-1121 Budapest, Konkoly Thege Mikl\'os \'ut 15--17, Hungary\\
$^{3}$ELTE E\"otv\"os Lor\'and University, Institute of Physics, P\'azm\'any P\'eter s\'et\'any 1/A, 1117 Budapest, Hungary}

\date{Accepted XXX. Received YYY; in original form ZZZ}

\pubyear{2021}

\begin{document}
\label{firstpage}
\maketitle

\begin{abstract}
OB associations, birthplaces of the most luminous stars, are key objects for understanding the formation of high-mass stars and their effects on their environments. The aim of this work is to explore the structure and kinematics of the Cepheus~OB2 association and characterize the history of star formation in the region -- in particular, the role of the Cepheus Bubble, surrounding Cepheus~OB2. Based on \gaia\ DR3 data we study the spatial and age distribution and kinematics of young stars in the region. We select candidate pre-main-sequence stars in the $M_\mathrm{G}$ vs. \gbp$-$\grp\  colour--magnitude diagram, and using a clustering algorithm, we identify 13 stellar groups belonging to Cep~OB2. Four groups, consisting of 10--13\,Myr old low- and intermediate-mass stars, are located in the interior of the bubble, and are part of the oldest subsystem of the association Cep~OB2a. Younger groups are found on the periphery. The tangential velocities suggest that some groups on the periphery were born in an expanding system of star-forming clouds, whereas others have been formed due to the collision of their parent cloud with the expanding bubble.
 
\end{abstract}




\section{Introduction}
OB~associations are unbound, dispersing groups of young, high luminosity stars, birthplaces of the high-mass stars of our Galaxy \citep{Blaauw1964,deZeeuw1999,Wright2020}. High-mass stars have a strong impact on the structure, physics and chemistry of the interstellar medium, thus their astrophysical importance is enormous. Studying the properties of associations is important for understanding their formation and evolution. Dimensions of OB~associations are 10--100 parsecs, and star formation in them may proceed for tens of million years. Most of them consist of substructures of different ages, including clusters of higher densities. Substructures of different ages are spatially and kinematically separated. Several associations are surrounded by supershells, created by the interaction of expanding ionization fronts, stellar winds, and supernova explosions of short-lived, high-mass stars with the surrounding interstellar medium \citep[e.g.][]{BGK1980}. Due to their 100--1000\,pc dimensions supershells are able to propagate star formation across the interstellar medium on very large scales \citep{McRay1987}.

The new astrometric and photometric data from the \gaia\ space telescope \citep{Gaia2016b} can answer several long-standing questions related to the structure, origin, and evolution of OB associations. These questions include the expansion of associations and propagating star formation. The low space density of stars in association ($< 0.1$\msun\,pc$^{-3}$) could reasonably be explained by the expansion of originally dense, compact systems.
\gaia\ data have shown that not all of the observed structure can be explained by expansion \citep[e.~g.][]{Wright2016,Wright2018,Kounkel2019}. \citet{Melnik2020} found expansion in five of the 28 examined associations. Kinematic studies by \citet{Lim2019,Lim2021,Lim2022} suggest that the formation of OB~associations may result from structure formation driven by supersonic turbulence, rather than from the dynamical evolution of individual embedded clusters. The observed age differences of substructures can be explained by star formation in the gas   compressed by the expanding ionization front and stellar wind bubble or supernova shock. 
 
Supershells may trigger star formation via several mechanisms: the expanding shock front may compress pre-existing clouds, or accumulate the ambient, low-density gas into a thin, dense layer, or may collide with ambient molecular clouds and with other bubbles \citep{EP2002}. These different mechanisms may result in new stellar populations which differ from each other in structure, velocity compared to the energy source, and star-forming time scales. Supershells, associated with nearby OB associations give an opportunity to study in detail various scenarios of propagating star formation. Precise parallaxes and proper motions, available in the \gaia\ data, allow us to study the spatial and kinematic substructures of OB associations, and explore the role of various types of triggered star formation. 

The subject of the present work is the Cepheus~OB2 association, discovered by \citet{Ambartsumian1949}. The luminous stars defining the association occupy an area of some $10\degr\times10\degr$ around the Galactic position $(l,b)=(102.1,+4.6)$  \citep*{Kun2008cep}. At a mean distance of 900\,pc \citep{Contreras2002} this angular size corresponds to a diameter of some 150\,pc. \citet{Simonson1968} identified 74 members of Cep~OB2 based on spectroscopy and \textit{UBV\/} photometry. Further members were identified based on \textit{HIPPARCOS} data \citep{deZeeuw1999}. Binary frequency among the high-mass stars of Cep~OB2 was examined by \citet{Peter2012}. They established that the multiplicity of massive ($M \ge 10$\msun) stars seems to be significantly higher than that of intermediate-mass stars. Two open clusters, the $\sim$4\,Myr old Trumpler~37, embedded in the \ion{H}{ii} region IC\,1396, and the $\sim12$\,Myr old NGC\,7160 are the historical subsystems of Cep~OB2 \citep{SA2005,SA2006}. Trumpler~37 itself is composed of several subgroups of different ages and structure, indicative of star formation triggered by interactions of the central O-type star HD\,206267 with ambient clumps of molecular gas \citep[e.g.][]{SA2015}.
 
Cep~OB2 is associated with the \textit{Cepheus Bubble\/}, a supershell of some 10\degr\ in angular diameter. It was identified in the \textit{IRAS\/} 60 and 100\,\micron\ images by \citet*{Kun1987}. \citet*{Abraham2000} 
studied the structure and kinematics of the bubble based on the data of the Leiden/Dwingeloo neutral hydrogen survey. The \ion{H}{i} data revealed the expansion of the bubble. Observations of the region in the 2.6-mm CO line by \citet{Patel1998} revealed an expanding shell of some 120\,pc in diameter, and containing $\sim4\times10^{5}$\msun\ gas. The morphology and kinematics suggested that the bubble was created by the stellar winds of the first generation of high-mass stars of Cep~OB2a during their lifetime of 8--10\,Myr, and accelerated by a supernova explosion $\sim1.7$\,Myr ago. The star-forming regions along the periphery of the bubble, such as Sh~2-129, IC\,1396, Sh~2-140, L1188, were probably formed a few million years after the first generation, by the fragmentation and collapse of the gas, compressed by the expanding stellar wind bubble. The molecular clouds associated with the bubble are birthplaces of the third generation of Cep~OB2 \citep[see][]{Szegedi2019}.

We study the \gaia~DR3 \citep{GaiaDR3} data of the stars in the area of the Cepheus~OB2 and the Cepheus Bubble. Our goal is to validate and extend the membership list of Cep~OB2, separate stellar populations of various ages and velocities, establish their relation to the expanding bubble, and find conclusions on the large-scale structure of Cep~OB2 and on the history of star formation.  We define our initial data set and the method of cluster selection in Sect.~\ref{sect:sample}. The tools applied to characterize the clusters are described in Sect.~\ref{sect:tools}. Our results are described in Sect.~\ref{sect:results}, discussed in Sect.~\ref{sect:view}, and briefly summarized in Sect.~\ref{Sect:sum}.

\section{Cep~OB2 membership based on \gaia}\label{sect:sample}

\subsection{High-mass members of Cep~OB2}\label{sect:hms}

Formation mechanism and environment of high-mass stars above 10\msun\ may differ from those of the lower mass stars \citep[e.g.][]{Tan2014}. The lists of luminous members of Cep~OB2, published by \citet{Simonson1968}, \citet{BH1989}, and \citet{deZeeuw1999} contain 90 stars above 10\msun. In order to compare the distribution and kinematics of high- and low-mass stars we compiled our list of massive stars from these (overlapping) tables, and searched for their \gaia~DR3 counterparts within 1\arcsec. In \gaia~DR3, 52 historical members, supergiants and main-sequence stars earlier than B3 have distances between 800 and 1000\,pc. Out of the 52, 8 stars have RUWE $>$ 1.4, indicating some issue with their astrometry, which could be caused by binarity \citep{lindegren-ruwe}. Most of them are indeed known binary or multiple systems \citep[e.g. BD+62\degr2078, HD\,204827, HD\,209744, HD\,239743,][]{Fabricius2002,Peter2012}. Surface distribution and tangential velocities of these high-mass stars are displayed in Fig.~\ref{fig:groups}, and listed in Table~\ref{tab:hms}. Several historical high-mass association members appear foreground (HD\,239712, HD\,199661) or background (HD\,235618, HD\,239758, HIP\,109603, BD+53\degr2387, HD 235783, HD\,235795, HD\,235813, HD\,240010, BD+57\degr2615, BD+53\degr2784, $\mu$~Cep, HD\,239978,	HIP\,111972) stars in \gaia~DR3. 

\begin{table*}
    \centering
    \caption{Sample of the high-mass stars described in Sect.~\ref{sect:hms}. The full table is available as a supplementary material. }
    \label{tab:hms}
    \begin{tabular}{l@{\hskip2mm}c@{\hskip2mm}cccccc@{\hskip2mm}c@{\hskip2mm}c}
        \hline
        \noalign{\smallskip}
        Name & Spectral type & Gaia DR3 source\_id & Distance & $\mu_{\alpha}^\star$ & $\mu_{\delta}$ & $\mathrm{RV_{Gaia}}$ & $\mathrm{RV_{lit}}$ & Memb & $\mathrm{RV_{ref}}$\\
        & & & (pc) & \multicolumn{2}{c}{(\masy)} & \multicolumn{2}{c}{(\kms)} & & \\
        \noalign{\smallskip}
        \hline
HD 198895 & B1Ve & 2183107416725856640 & $864_{-11}^{+12}$ & $-2.72\pm0.02$ & $-4.91\pm0.02$ & $\cdots$ & $-23.0\pm7.4$ & 1 & 7\\
HD 199308 & B1.5V & 2189911847512999168 & $811_{-13}^{+16}$ & $-5.21\pm0.03$ & $-4.68\pm0.02$ & $\cdots$ & $-23.0\pm4.5$ & 1 & 6\\
HD 200857 & B3III & 2188891844319403648 & $849_{-12}^{+16}$ & $-2.99\pm0.02$ & $-4.70\pm0.02$ & $\cdots$ & $-14.0\pm4.4$ & 1 & 6\\
HD 204150 & B2III & 2191963398772774144 & $863_{-16}^{+20}$ & $-2.18\pm0.03$ & $-2.98\pm0.03$ & $\cdots$ & $-18.0\pm12.9$ & 1;3 & 7\\
HD 205139 & B1Ib & 2191787957952122752 & $855_{-28}^{+32}$ & $-1.80\pm0.06$ & $-3.75\pm0.05$ & $\cdots$ & $-14.5\pm2.9$ & 1;3 & 6\\
        \hline
    \end{tabular}
\smallskip
\flushleft{\small Membership references: 1 - \citet{Simonson1968}, 2 - \citet{BH1989}, 3 - \cite{deZeeuw1999}.\\
 RV references:  1 - \citet{Wilson1953}, 2 - \citet{Petrie1961}, 3 - \citet{Hilditch1982}, 4 - \citet{Barbier-Brossat2000}, 5 - \citet{Pourbaix2004}, 6 - \citet{Gontcharov2006}, 7 - \citet{Kharchenko2007}, 8 - \citet{Boyajian2007}, 9 - \citet{deBruijne2012}, 10 - \citet{Holgado2018}, 11 - \citet{Katz2022}} 
\end{table*}

\subsection{Members defined by distances and tangential velocities}
\subsubsection{\gaia~sample}\label{sect:gaiasample}
We selected all sources from \gaia~DR3 database with:
\begin{enumerate}[align=left]
    \item $96\degr < l < 108\degr$,
    \item $2\degr < b < 12\degr$,
    \item $800 < d < 1000\,\mathrm{pc}$,
    \item $\varpi/\sigma_\varpi \geq 10$,
    \item $|\mu_{\alpha}^\star/\sigma_{\mu_{\alpha}^\star}| \geq 5$,
    \item $|\mu_{\delta}/\sigma_{\mu_{\delta}}| \geq 5$,
    \item $\mathrm{RUWE} \leq 1.6$,
\end{enumerate}
where $l$ and $b$ are the galactic longitude and latitude, $d$ is the distance from \citet{BJ2021}, $\varpi$, $\mu_{\alpha}^\star$, $\mu_{\delta}$ and $\sigma_\varpi$, $\sigma_{\mu_{\alpha}^\star}$, $\sigma_{\mu_{\delta}}$ are the parallax, proper motion in right ascension and declination and their uncertainities, respectively. 

We selected the pre-main-sequence stars from the \gaia\ sample using their $M_\mathrm{G}$ vs. \gbp$-$\grp\ colour-magnitude diagram, corrected for interstellar extinction (Fig.~\ref{fig:cmd_ini}). We corrected the \gaia\ colour indices and magnitudes of each star using the Python implementation\footnote{\url{https://github.com/edober/dust_maps_3d}} of the 3D dust maps  STILISM \citep{Lallement2018}. STILISM gives the \ebv\ colour excess as a function of  galactic coordinates and distance. We transformed the extinction into the \gaia\ bands with the coefficients in table~3. of \citet{Wang2019}, assuming $R_\mathrm{V} = 3.1$.

Following the method described in \citet{Zari2019} we have used the 10-Myr isochrone from PARSEC \citep{Bressan2012} to define an area in the de-reddened $M_\mathrm {G}$ vs. \gbp$-$\grp\ colour--magnitude diagram, occupied by young stellar objects (YSOs). Figure~\ref{fig:cmd_ini} suggests that the stellar sample between the dashed lines is contaminated with main-sequence stars, located in the 800--1000\,pc distance interval but not related to the association. To tighten the selection we searched for spatially and kinematically coherent groups in the sample selected from the colour--magnitude diagram.

\begin{figure*}
    \centering
    \includegraphics[width=\textwidth]{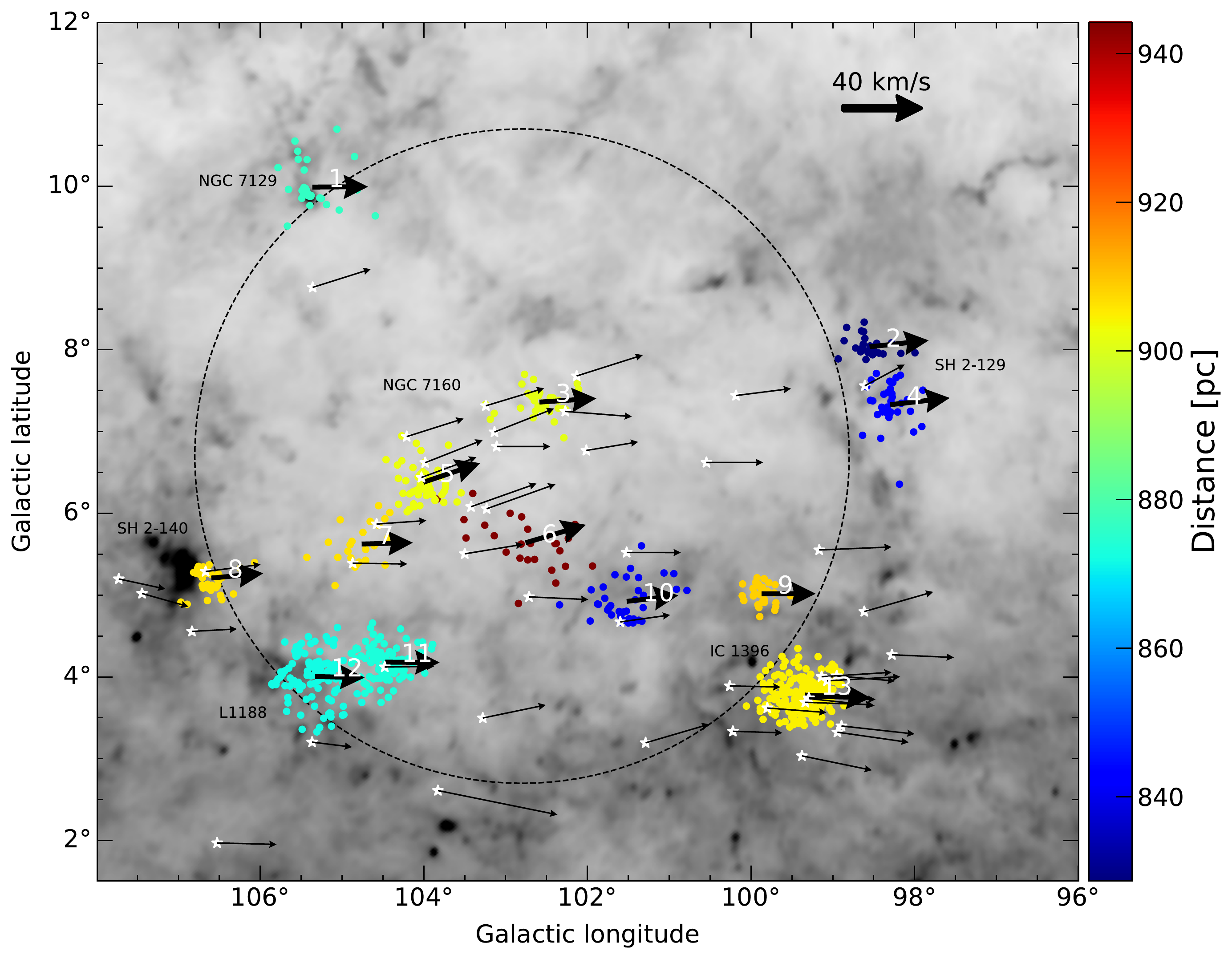}
    \caption{Members of Cepheus~OB2 overplotted on the \planck\ 857 GHz map of the region. Circles show the members of the stellar groups found by HDBSCAN, and the white star symbols represent the high-mass ($M >10$\msun) members of the Cepheus~OB2 association from Table~\ref{tab:hms}. The arrows show the textbf{mean} tangential velocities of the groups and the OB stars, compared to the \lsr, while the colour bar shows the textbf{mean} distance of each group. The circle indicates the rough size and position of the Cepheus Bubble.}
    \label{fig:groups}
\end{figure*}

\begin{figure}
    \centering
    \includegraphics[width=\columnwidth]{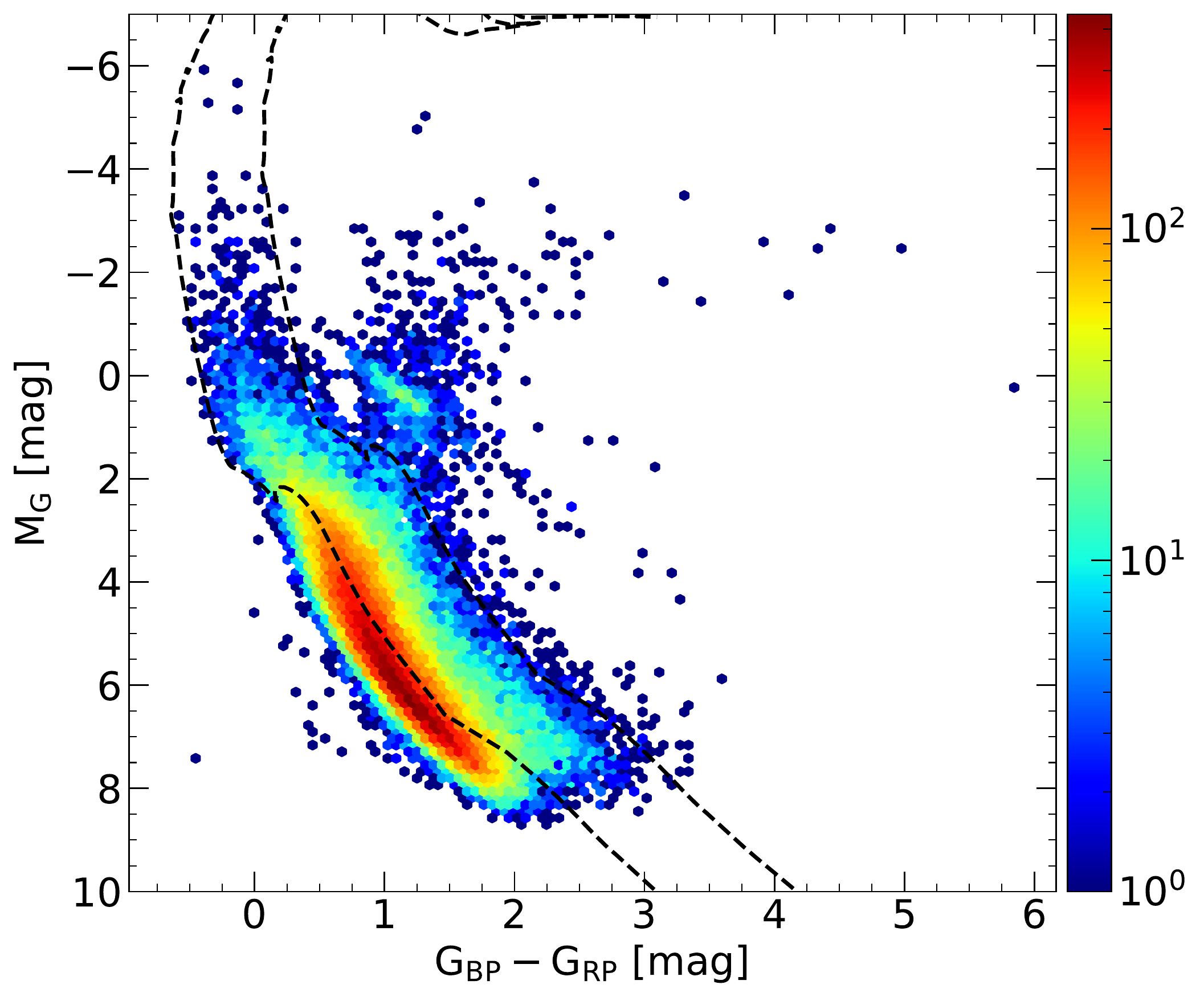}
    \caption{De-reddened colour--magnitude diagram of the sources described in Section~\ref{sect:sample}. The dashed lines from \citet{Zari2019} border the area defined to select the pre-main-sequence stars.}
    \label{fig:cmd_ini}
\end{figure}

\subsubsection{Search for clusters}\label{sect:cls}

We used the Python-implemented package of Hierarchical Density-Based Spatial Clustering of Applications with Noise (HDBSCAN) clustering algorithm \citep{mcinnes2017hdbscan} to find overdensities in our sample obtained in Sect.~\ref{sect:gaiasample}. An advantage of this clustering algorithm over other methods is that HDBSCAN can identify groups with various densities and arbitrary shapes. Main parameters of HDBSCAN are \mincls, \mins\ and \csm. $\mathtt{Minimum\_cluster\_size}$ defines the minimal number of data points a cluster must hold at least, while the \mins\ defines how conservative a clustering is: the bigger the \mins\ is, the more points are considered as noise. At default, \csm\ uses the Excess of Mass (EOM) approach to select one or two large and several smaller clusters. For more homogeneous, smaller clusters, we can use the Leaf method. For a detailed description of the algorithm, see the  website\footnote{\url{https://hdbscan.readthedocs.io/en/latest/index.html}} of the package. 

There are multiple ways for detecting stellar clusters in the \gaia\ data: some used ICRS coordinates, parallaxes and proper motions \citep[e.g.][]{Kounkel2019}, some used heliocentric \textit{XYZ} coordinates and tangential velocities multiplied by a constant \citep[e.g.][]{Kerr2021}. Using the distance of 900\,pc from \citet{Contreras2002} we transformed the Galactic coordinates $l$ and $b$ of the stars into 2D cartesian coordinates compared to the centre of the studied field. We also calculated the \vllsr\ and \vblsr\ Galactic tangential velocity components compared to the Local Standard of Rest (\lsr) using the $(UVW)_{\sun}$ values from \citet{Schonrich2010}. The fifth dimension is the distance. Due to the different units we standardized the data in each dimension by subtracting the mean and dividing with the largest standard deviation of the dimensions that share the same unit, which are the distance and \vllsr\, respectively. We adopted $\mathtt{minimum\_cluster\_size}=25$ and $\mathtt{minimum\_samples}=25$ with the Leaf method.

\section{Characterization of the groups}\label{sect:tools}
HDBSCAN found originally 10 groups, consisting altogether of 874 stars. The final Groups 2 and 4, 5 and 7, and 11 and 12 were originally merged into three larger groups, respectively, but their spatial distribution suggested that each of the three can be split into two smaller subgroups. The Groups~2 and 4 pairs we split at $b=7.8^\circ$. For dividing the Group~5--7 and 11--12 pairs we used two-component Gaussian mixture model from \texttt{scikit-learn} \citep{scikit-learn} to fit to their spatial distributions. These actions resulted in 13 groups. We regard these 13 groups as substructures of Cep~OB2, and examine how their properties reflect the history of star formation in the region.  
Figure~\ref{fig:groups} shows the distribution of the clustered sources in Galactic coordinates, overplotted on the \planck\ 857\,GHz image of the region. The mean tangential velocities of the groups, compared to the \lsr, are also indicated. The colouring shows the mean distances of groups. The mean coordinates, distances and velocities of the groups are shown in Table~\ref{tab:groups}. Known clusters and clouds from the literature, associated with the individual groups, are indicated. Surface distribution of stars in individual groups, along with their tangential velocity vectors compared to the mean velocity of the groups listed in Table~\ref{tab:groups}, are displayed in Fig.~\ref{fig:group1}a and Figs.~\ref{fig:group2}a--\ref{fig:group13}a. Distance histograms and \vllsr\ vs. \vblsr\ diagrams plotted in Figs.~\ref{fig:group1}b and \ref{fig:group2}bc--\ref{fig:group13}bc. Table~\ref{tab:groupmembers}, containing a detailed list of group members, is available in machine-readable form.  

\subsection{Radial velocities}\label{sect:rv}
Gaia DR3 provides radial velocities for stars with $G_{\mathrm{RVS}}<14$\,mag \citep{Katz2022}. From the 874 group member stars, 191 have radial velocites measured by \gaia\/. The number of the stars with available radial velocities for each group are provided in Table~\ref{tab:groups}.

\subsection{Variable stars}\label{sect:vari}
\begin{figure}
    \centering
    \includegraphics[width=\columnwidth]{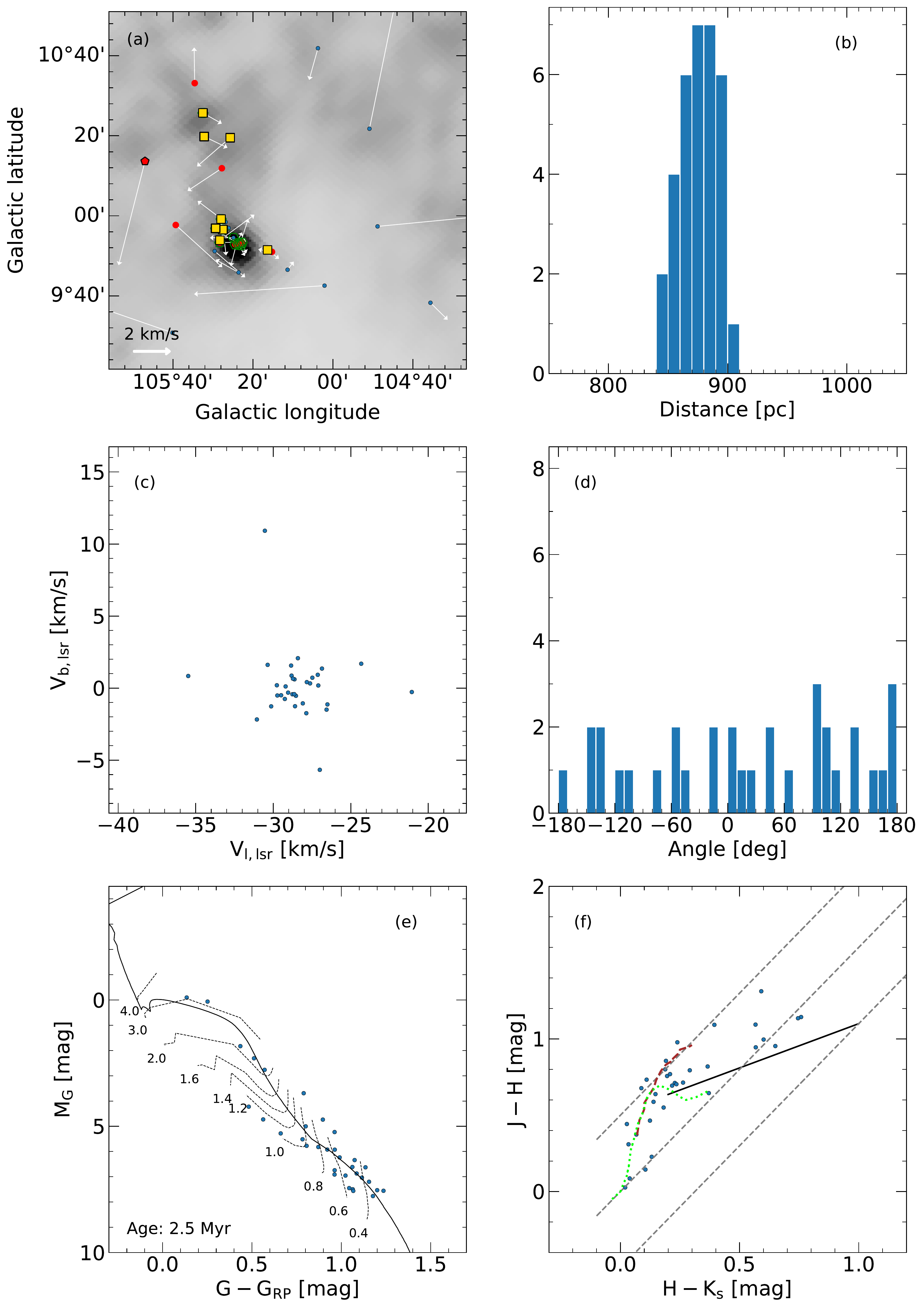}
    \caption{\textsl{a}: Distribution of the members of Group~1, plotted on the \planck\ 857\,GHz image. The arrows indicate the tangential velocities of the stars compared to the mean tangential velocity of the group. Gold squares and the red pentagon indicate the Class~II and Class~I sources classified by the \wise\ colour indices, respectively. YSO-candidates from \citet{Marton2022} are overplotted with red circles. YSOs from \citet{Dahm2015} and \citet{Kun2009} are overplotted with green hexagons. \textsl{b}: Histogram of distances. \textsl{c}: Tangential velocity components compared to the \lsr. \textsl{d}: Distribution of the angles between the radial vector of each star from the group centroid and relative tangential velocities (see \textsl{a}). \textsl{e}: De-reddened $M_\mathrm{G}$ vs. \gbp--\grp\ colour--absolute magnitude diagram. The solid line shows the best-fitted PARSEC isochrone. Dashed lines represent evolutionary tracks from 1 to 20 Myrs of several stellar masses. For 1.4\msun\ and below we plotted CIFIST isochrones \citep{Baraffe2015}. \textsl{f}: De-reddened \tmass\ colour--colour-diagram showing sources with quality criteria 'AAA'. The plots of the other groups are presented in the Appendix.}
    \label{fig:group1}
\end{figure}

Gaia DR3 provides a list containing 24 types of variable stars identified with machine learning methods \citep[see][for details]{Rimoldini2022,Eyer2022}. We crossmatched our group members with it, and found that 355, $\sim41\,\%$ of the member stars are classified as YSO-candidates in \citet{Marton2022}. These stars are overplotted with red circles in Fig.~\ref{fig:group1}a and Figs.~\ref{fig:group2}a--\ref{fig:group13}a. Furthermore 60 stars were identified as RS Canum Venaticorum type variable stars. Additional 17 stars were classified as either eclipsing binaries \citep{Mowlavi2022}, solar-like variables, $\alpha^2$~CVn/magnetic chemical peculiar star/rapidly oscillating Am/Ap~star/SX~Ari variable stars \citep{Distefano2022} or $\delta$~Sct/$\gamma$~Dor/SX~Phe. Column $\mathtt{best\_class\_name}$ in Table~\ref{tab:groupmembers} contains these information.

\subsection{Colour--absolute magnitude diagrams}\label{sect:cmd}
Figures\,\ref{fig:group1}e and \ref{fig:group2}e--\ref{fig:group13}e show the \textit{M}$_\mathrm{G}$ vs. \g$-$\grp\ de-reddened colour--absolute magnitude diagrams (CMD) of each group. We downloaded PARSEC isochrones \citep{Bressan2012} with logarithmic ages from 6 to 7.5 with a step of 0.1. The CMDs show the best-fitted isochrones, as well as  evolutionary tracks for several initial masses. For 1.4\msun\ and below we plotted CIFIST \citep{Baraffe2015} tracks.

\subsection{2MASS colour--colour diagrams}\label{sect:2mass}
We crossmatched our data with the \tmass\ \citep{2mass} database. We transformed ICRS coordinates of the stars from \gaia's J2016 epoch into J2000 epoch to search for \tmass\ counterparts. Then we searched for coinciding \tmass\ sources within 1\,arcsec. We calculated the extinctions in the $JHK_\mathrm{s}$ bands of \tmass\ by the method described in Sect.~\ref{sect:sample}. De-reddened \jb$-$\h\ vs. \h$-$\ks\ colour--colour diagrams (CCD) are displayed in Figs.~\ref{fig:group1}f and \ref{fig:group2}f--\ref{fig:group13}f, showing only the stars with photometric quality criteria 'AAA'. 

\subsection{\wise\ colour--colour diagram}\label{sect:wise}
\begin{figure}
    \centering
    \includegraphics[width=\columnwidth]{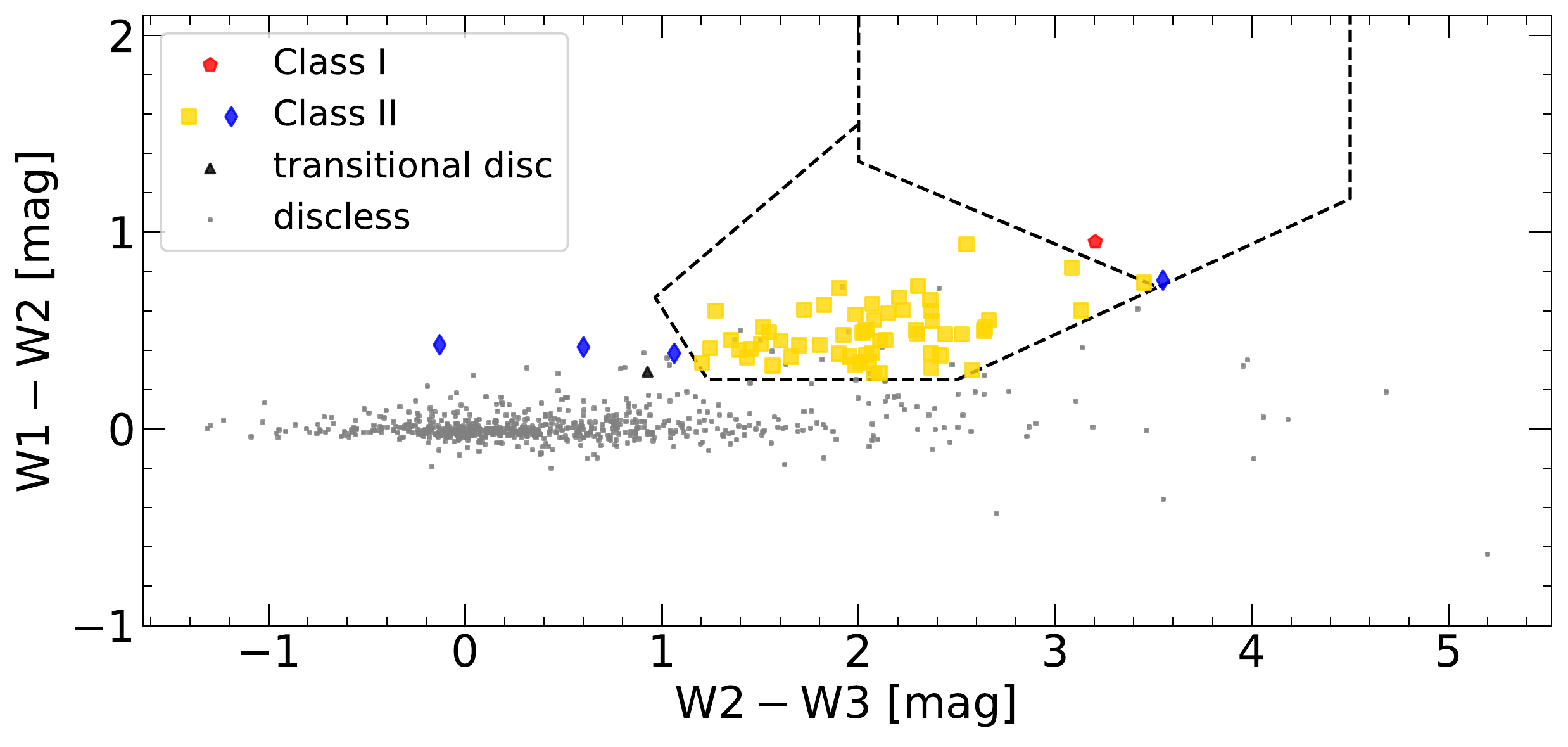}
    \caption{AllWISE colour--colour diagram of the group members. The dashed lines border the areas of Class I and II stars, according to \citet{Koenig2014}.}
    \label{fig:allwise}
\end{figure}

We also searched for counterparts in the \textit{AllWISE\/} \citep{allwise} catalogue to identify potential disc-bearing stars. We found counterparts of 733 stars within 1\,arcsec. We followed the methods described by \citet{Koenig2014} to find disc-bearing stars. From these \textit{AllWISE\/} sources, 316 fulfil the quality criteria essential for constructing their $W1-W2$ vs. $W2-W3$ colour--colour diagram, displayed in Fig.~\ref{fig:allwise}. One star is classified as a Class~I source, 58 of the sources are identified as Class~II sources and one star is classified as transitional disc bearing star. Supplemented with \tmass\ data we identified 4 more Class~II sources. These stars are plotted as red pentagons, gold squares, black triangles and blue diamonds, respectively in Figs.~\ref{fig:group1}a and \ref{fig:group2}a--\ref{fig:group13}a.

\subsection{Internal motions of the groups}\label{sect:vint}
We examined the internal motions of the substructures of Cep~OB2. Following the method applied by \citet{Lim2019} we determined the angle between the radial vector of each star from the centre position of the group and its relative velocity vector. Figures~\ref{fig:group1}d and \ref{fig:group2}d--\ref{fig:group13}d show the histograms of these angles. Clustering of the angles around 0\degr\ is indicative of radial expansion of the group, whereas peaks around $\pm180\degr$ suggest contraction. 

\begin{landscape}
\begin{table}
    \centering
    \caption{Mean coordinates, distances, velocities, \ebv\ colour excesses and ages for each stellar group. The last five columns show the number of group member stars, stars with measured radial velocities, identified as YSO \citep{Marton2022}, RS Canum Venaticorum variable, and other type of variables, listed in Sect.~\ref{sect:vari}. The table is available as supplementary material.}
    \label{tab:groups}
    \begin{tabular}{clccccccccccccc}
        \hline
        \noalign{\smallskip}
        Group & Associated object &	$\mathrm{l_{mean}}$ & $\mathrm{b_{mean}}$ & $\mathrm{d_{mean}}$ &  $\mathrm{v_{l,LSR,mean}}$ & $\mathrm{v_{b,LSR,mean}}$ & \ebv & age & N$_\star$ & $\mathrm{N_{rad}}$ & $\mathrm{N_{YSO}}$ & $\mathrm{N_{RS}}$ & $\mathrm{N_{other}}$\\
        & & \multicolumn{2}{c}{(deg)} & \multicolumn{1}{c}{(pc)} & \multicolumn{2}{c}{(\kms)} & (mag) & (Myr) & & & & \\
       \noalign{\smallskip} 
       \hline
1 & NGC\,7129 & 105.36 & 9.99 & $876\pm15$ & $-28.4\pm2.2$ & $0.2\pm2.4$ & 0.21 & 2.5 & 33 & 6 & 16 & 3 & 1\\
2 & SH~2-129, UBC\,385 & 98.55 & 8.04 & $829\pm13$ & $-30.1\pm1.7$ & $3.3\pm1.3$ & 0.49 & 10.0 & 27 & 12 & 5 & 2 & 1\\
3 & UBC\,10a, UBC\,167 & 102.59 & 7.36 & $901\pm6$ & $-29.0\pm0.9$ & $1.9\pm0.5$ & 0.43 & 10.0 & 33 & 13 & 3 & 7 & 1\\
4 & $\cdots$ & 98.30 & 7.33 & $842\pm13$ & $-30.3\pm1.2$ & $3.4\pm1.2$ & 0.48 & 15.8 & 33 & 11 & 7 & 3 & 1\\
5 & NGC\,7160 & 104.01 & 6.38 & $902\pm11$ & $-28.9\pm1.0$ & $9.8\pm3.0$ & 0.29 & 10.0 & 64 & 19 & 7 & 7 & 4\\
6 & UBC\,10b & 102.76 & 5.64 & $944\pm7$ & $-30.8\pm1.1$ & $9.2\pm1.3$ & 0.34 & 10.0 & 26 & 8 & 1 & 2 & 0\\
7 & [KPR2005]~117 & 104.76 & 5.63 & $906\pm4$ & $-26.1\pm1.0$ & $0.7\pm0.9$ & 0.32 & 10.0 & 22 & 6 & 5 & 3 & 0\\
8 & SH 2-140, Pismis--Moreno~1 & 106.60 & 5.21 & $906\pm5$ & $-26.4\pm3.1$ & $2.4\pm1.8$ & 0.61 & 10.0 & 45 & 11 & 19 & 1 & 0\\
9 & IC\,1396, L1116 & 99.87 & 5.02 & $908\pm9$ & $-27.6\pm1.2$ & $0.1\pm1.1$ & 0.32 & 5.0 & 30 & 9 & 13 & 3 & 0\\
10 & UPK\,169, Theia 131 & 101.52 & 4.92 & $841\pm7$ & $-26.2\pm1.6$ & $3.0\pm1.9$ & 0.27 & 4.0 & 39 & 15 & 5 & 0 & 0\\
11 & Alessi--Teutsch~5 & 104.47 & 4.18 & $873\pm14$ & $-27.8\pm1.9$ & $-0.2\pm1.4$ & 0.37 & 7.9 & 93 & 22 & 21 & 6 & 1\\
12 & L1188, [BDS2003]~30 & 105.33 & 4.00 & $872\pm22$ & $-25.2\pm1.8$ & $-0.5\pm1.7$ & 0.6 & 7.9 & 134 & 23 & 66 & 7 & 3\\
13 & IC\,1396, Trumpler\,37 & 99.34 & 3.78 & $904\pm24$ & $-33.5\pm1.9$ & $-1.5\pm1.6$ & 0.35 & 5.0 & 295 & 36 & 187 & 16 & 5\\
        \hline
    \end{tabular}
\end{table}
\end{landscape}

\section{The substructures of Cep~OB2}\label{sect:results} 
The historical substructures of Cep~OB2 are NGC\,7160, surrounded by a number of evolved high-mass stars (Cep OB2a), and IC\,1396 (Cep~OB2b) \citep{Simonson1976}. These represent two epochs of star formation in the volume of the association. The Cepheus Bubble has revealed new probable subgroups and suggested connection between them. Our search for stellar groups in the \gaia\ data resulted in further subgroups. Except Group 4 they coincide with known clusters, but most of them have not been coupled to Cep~OB2 in the literature. We describe in this section the \gaia\ view of the substructures. 

Most of clustered stars have \g\ < 18\,mag, only 33 of the 874 group members are fainter than this limit. The foreground \av\ extinctions are between 1--1.8\,mag. These data suggest that the masses of the detected group members are above 0.6 and 0.7\msun\ in a 5 and 10\,Myr old group, respectively. 

\paragraph*{Group~1,} the northernmost group, contains the well-known, compact young cluster NGC\,7129, and a few stars to the west and north of the cluster. \citet{Kun1987} associated NGC\,7129 with the Cepheus Bubble based on morphological considerations. \gaia\ data, in accordance with the VLBA parallax published in \citet{Reid2014} confirm that this cluster is located on the surface of the Cepheus Bubble. HDBSCAN identified 33 members, 8 of them are classified as Class~II stars, and 1 is identifies as a Class~I source. The colour--magnitude diagram (Fig.~\ref{fig:group1}e) suggests an age of 2.5~Myr. Fig.~\ref{fig:group1}f suggests significant intra-cluster extinction, which may affect the appearance of the colour--magnitude diagram. Eleven of the member stars appear in \citet{Dahm2015} and one in \citet{Kun2009} as YSOs. They are overplotted with green hexagons in Fig.~\ref{fig:group1}a.

\paragraph*{Groups~2 and 4} are located at the western edge of the Cepheus Bubble. According to the CMDs in Figs.~\ref{fig:group2}e and \ref{fig:group4}e they are 10 and 15.8\,Myr old, and most of their members are more massive than 1\msun. Unlike other groups, the detected members of Group~2 are aligned evenly along the fitted isochrone: the number of stars above 2\msun\ is nearly identical with that in the 1\,M$_{\sun} <$ M $<2\,$M$_{\sun}$ interval. Group~2 is the nearest group with a mean distance of $829\pm13$\,pc. It coincides with the cluster UBC\,385, identified by \citet{Castro-Ginard2020}, and contains several B8--A0 type stars, classified as Cep~OB2 members by \citet{Alksnis1958}. It is the central cluster of the ring-shaped \ion{H}{ii} region Sh~2-129, excited by the O9.5\,IV type \citep{Sota2014} component of the young triple system HD\,202214 \citep{Balega2004}. Though the distance of HD\,202214, projected at the cluster centre is uncertain, its spectral type and brightness suggest its membership. None of the group members were classified as disc-bearing stars. The neighbouring Group~4 was merged with Group~2 by HDBSCAN, but a closer inspection has shown that they differ slightly in distance.

\paragraph*{Groups~3, 5, 6 and 7} are projected inside the Cepheus Bubble. The colour--magnitude diagrams suggest ages of 10~Myr, thus these groups belong to the first generation of Cep~OB2 \citep{SA2005}. Groups~3, 6, and 7 coincide with the known clusters UBC\,10a, UBC\,10b \citep{Castro-Ginard2018}, and [KPR2005]~117 \citep{Kharchenko2005}, respectively. The groups are projected close to each other, but differ in distance, space velocity and mass distribution. Compared to the neighbouring groups Groups~5 and 6 have large latitudal tangential velocity components, and high proportion of low-mass ($M \leq 1$\msun) members. The angular distribution of the tangential velocity vectors in Fig.~\ref{fig:group5}d shows the expansion of Groups~5. These groups lack interstellar matter. The dark cloud L1178 is located at the south-eastern edge of the region, but the extinction towards the cloud rises around 700\,pc, suggesting that they are not related.

\paragraph*{Group~8} is associated with the cluster Pismis--Moreno~1 \citep*{Pismis1979} that contains HD\,211880, the B0.5V type exciting star of SH~2-140. The colour--magnitude diagram indicates an age around 10~Myr, in accordance with earlier estimates \citep[e.g.][]{Cantat2020b}. Four stars were designated as Class~II star by the method of \citet{Koenig2014}. Star formation is on-going in the molecular cloud bordered by the \ion{H}{ii} region \citep{Gutermuth2009}. Quite a few Class~II \spitzer\ sources, without reliable \gaia\ data, are projected within this group \citep[see fig.~1 in][]{Gutermuth2009}, indicative of either their membership or the overlapping of the older Pismis--Moreno~1 and the embedded S140 cluster. 

\paragraph*{Groups~9 and 13} are located at the IC\,1396 region. Several dark clouds of \citet{Lynds1962} are found in the region. The extinction towards L1086, L1096, L1102 and L1116 rises around 900\,pc, indicating that all these clouds may be related to IC\,1396. Group~13 corresponds to the cluster Trumpler~37, containing the exciting star of IC\,1396. We have identified 39 Class~II sources and a transitional disc-bearing star in Group~13. The 295 members include 69 low-mass YSOs identified by \citet{SA2005}, and the intermediate-mass YSO IRAS~21365+5713 from \citet{Contreras2002}. Figure~\ref{fig:group13}d indicates the expansion of Tr\,37. According to the CMDs, the ages of groups are around 5~Myrs.

\paragraph*{Group~10} corresponds to the cluster UPK\,169 \citep{Cantat2020a}. It also appears in \citet{Kounkel2019} as part of Theia~131. It is projected halfway between NGC\,7160 and Tr~37, but its distance is some 60\,pc smaller. The mean age of the group members is about 4~Myr. Based on \textit{AllWISE\/} data none of the 39 members have infrared excess, characteristic of YSO discs. The brightest member of the group is HD\,207538, an O9.7IV spectral-type star \citep{Sota2011}. 
 
\paragraph*{Groups~11 and 12} correspond to the known clusters Alessi--Teutsch~5 and [BDS2003]~30, respectively (\citealp{Kharchenko2005, BDS2003}). Eleven stars of Group~12 were identified as Class~II sources based on \textit{AllWISE\/} and \tmass\ data. The most luminous member of Group~11 is the B1IV type component of the eclipsing binary V*\,V446\,Cep. The star-forming dark cloud L1188 is associated with Group~12. Ten of Group~12 members were classified as young stars in \citet{Szegedi2019}. The CMDs suggest ages of 7.9~Myr for both groups.

\begin{table}
    \begin{center}
    \caption{Column description of the table containing the members of the stellar groups. The list is available as supplementary material.}
    \label{tab:groupmembers}
    \begin{tabular}{ll}
        \hline	    
        Column & Description \\
        \hline
        group & Group identifier \\ 
        name & Star identifier \\
        source\_id & Gaia DR3 source id \\
        TMASS & 2MASS identifier \\
        AllWISE & AllWISE identifier\\
        ra & RA at J2016 \\
        dec & DEC at J2016 \\
        l & Galactic longitude \\
        b & Galactic latitude \\
        d & Distance from \citet{BJ2021} \\
        dmin & 16th percentile of distance posterior\\
        & from \citet{BJ2021} \\
        dmax & 84th percentile of distance posterior\\
        & from \citet{BJ2021} \\
        parallax & Parallax \\
        parallax\_error & Error of the parallax \\
        pmra & Proper motion in right ascension \\
        pmra\_error & Error of the proper motion in RA \\
        pmdec & Proper motion in declination \\
        pmdec\_error & Error of the proper motion in DEC \\
        radial\_velocity & Radial velocity\\
        radial\_velocity\_error & Error of the radial velocity\\
        ruwe & Renormalized unit-weight error \\
        phot\_g\_mean\_mag & Magnitude in \g\ band \\
        phot\_bp\_mean\_mag & Magnitude in \gbp\ band \\
        phot\_rp\_mean\_mag & Magnitude in \grp\ band \\
        Jmag & Magnitude in 2MASS \jb\ band \\
        Hmag & Magnitude in 2MASS \h\ band \\
        Kmag & Magnitude in 2MASS \ks\ band \\
        eb\_v & \ebv\ from STILISM\\
        allwise\_class & YSO type by the classification of\\
        & \citet{Koenig2014}\\
        best\_class\_name & Variable star type appear in\\
        & gaiadr3.vari\_classifier\_result\\
        & \citep{Rimoldini2022}\\
        \hline
    \end{tabular}
    \end{center}
\end{table}

\section{Overall view of Cepheus~OB2}\label{sect:view}
\begin{figure*}
    \centering
    \includegraphics[width=\textwidth]{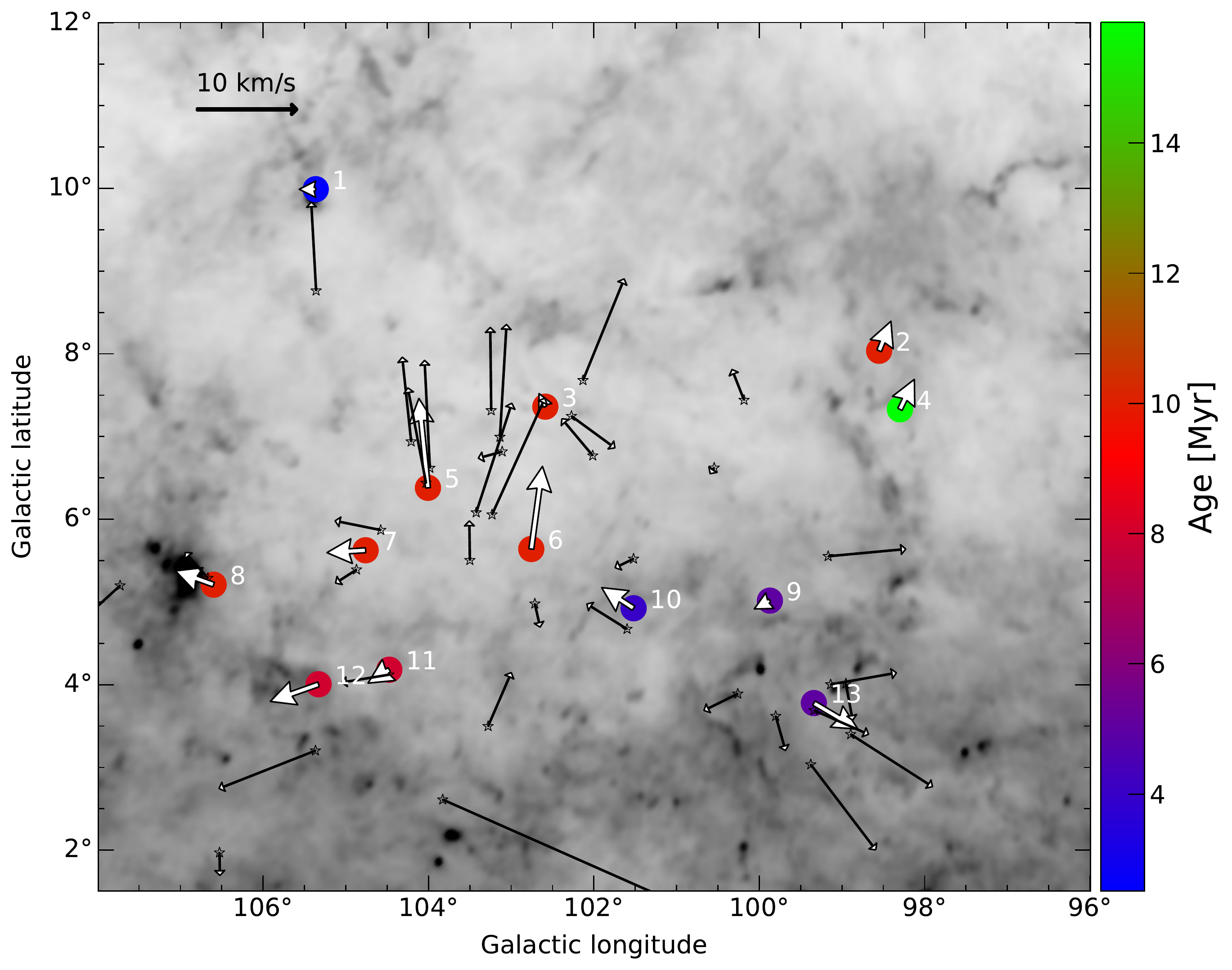}
    \caption{2D structure of Cepheus~OB2. Large circles and smaller star symbols represent the groups from Table~\ref{tab:groups} and the high-mass stars from Table~\ref{tab:hms}, respectively. Arrows indicate the mean tangential velocities, compared to the mean motion of the system, and the colour bar indicates group ages.}
    \label{fig:motion}
\end{figure*}

The line-of-sight dimension of the system of groups is some 120\,pc, similar to the apparent diameter of the Cepheus Bubble. To get an insight into the internal motions of the stars associated with the Cepheus Bubble we calculated the mean motion from the velocity data in Table~\ref{tab:groups}. The mean velocities are $v_\mathrm{l}=-16.13\pm4.01$\,\kms and $v_\mathrm{b}=-5.43\pm3.87$\,\kms. In Fig.~\ref{fig:motion} we represent the tangential velocities of the groups compared to the mean values. The precision of available stellar radial velocities is not sufficient for studying the three-dimensional velocity structure.

The colour--magnitude diagrams show that star formation started some 10 million years ago in the studied region. Groups~3, 5, 6, 7 are located inside the volume bordered by the Cepheus Bubble, in the volume occupied by the evolved high-mass members of Cep~OB2a. These groups are new components of Cep~OB2a. The mean distance of this subsystem is 900\,pc,  the line-of-sight size is about 50\,pc, and the dispersion of the tangential velocities is some 10\,km\,s$^{-1}$. The tangential velocities indicate the expansion of the largest group, most probably due to the disappearance of the the parental cloud. Expansion of the whole Cep~OB2a cannot be demonstrated by the method applied in Sect.~\ref{sect:vint}. The spatial and velocity structure suggest that each group had its own parent cloud clump, forming stars independently. Most of the high-mass stars were not selected by HDBSCAN as group members, the only exception is the B1\,III type HD\,208218, the most massive member of Group~5. The reason may be that the velocities of the high-mass stars are influenced by their probable multiplicities. 

Groups~1, 2, 4, 8, 9, 12 and 13 line along the apparent perimeter of the Bubble. Except Groups~2, 4 and 8, these groups are younger than Cep~OB2a. The diversity of their stellar contents, ages, and velocities suggest a complex history of star formation. The expansion of the by now extinct \ion{H}{ii} region of Cep~OB2a might have affected the formation of these groups, and/or the evolution of circumstellar discs of the group members. 

Figure~\ref{fig:motion} shows that the Groups 11, 12, and 13 move radially outwards from the centre, suggesting that their parent clouds were accelerated by the shock wave from the expanding ionization front of the OB stars of Cep~OB2a. Each of these groups is younger than Cep~OB2a. Groups~12 and 13 are associated with large amounts of molecular gas \citep[][respectively]{Szegedi2019,SA2015}, sites of active star formation. The colour--magnitude diagrams suggest that Groups~11 and 12 are nearly coeval. The difference in the proportion of disc-bearing members  may result from the different environments \citep[cf.][]{Dale2013}. 

The small groups Group~2 (UBC\,385) and 4 are projected near the westernmost wall of the Cepheus Bubble. \gaia\ data have shown that these groups are some 70\,pc closer to us than the central groups. Their ages, suggested by the \gaia\ CMDs, are similar to that of Cep~OB2a. Group~2 is centred on the \ion{H}{ii} region S129. Since O-type stars and \ion{H}{ii} regions are generally younger than 10\,Myr \citep[e.g.][]{Tremblin2014}, we may speculate that the exciting star HD\,202214 was born later than the lower-mass members of the group. SPH simulations by \citet{Dale2013} show that such a situation may occur in molecular clouds near  ionizing stars. Precise astrometric data of the HD\,202214 triple system and more data on the cluster population may clarify the nature of star-forming processes in the region of S129, and its connection with the Cepheus Bubble.

Group~8 (Pismis--Moreno~1) is another 10~Myr old group associated with an \ion{H}{ii} region. The main-sequence lifetime  of the B0.5\,V type exciting star HD\,211880 is compatible with this age. The molecular cloud bordered by the ionization front contains the S140 embedded cluster \citep{Gutermuth2009}. This morphology suggests that probably Group~8 and HD\,211880 itself were formed in the same molecular cloud, and star formation propagates radially outwards from the central Cep\,OB2a. However, the role of the expanding Cepheus Bubble in the formation of this cluster is unlikely because of the apparently similar ages of Cep~OB2a and Group~8.

Group~9 is a small group at the outskirts of IC\,1396, associated with dark clouds, bordering the \ion{H}{ii} region. Its age is similar to that of Tr\,37, whereas their tangential velocities, and probably star formation histories are different. 

Group~10 \citep[UPK\,169,][]{Cantat2020a} is a $\sim4$\,Myr old small group, containing the O9\,V type star HD\,207538. It is projected inside the Bubble, and its distance of $841\pm7$\,pc suggests  association with the near wall. In spite of its young age, no disc-bearing low-mass star has been detected in this group, probably due to the disruptive radiation of the hot star. The parent cloud of this group was probably overrun by the expanding ionization front of Cep~OB2a. Similarly, collision of the expanding bubble with the southernmost edge of an ambient giant molecular cloud, located at Galactic latitudes 10--13\degr\ \citep{Grenier1989}, might have played a role in the formation of Group~1  (NGC\,7129).  

\section{Summary} \label{Sect:sum}
We have studied the stars from \gaia~DR3 in the region of the Cepheus~OB2 association between $96\degr < l < 108\degr$, $2\degr < b < 12\degr$ and $800 < d < 1000\,\mathrm{pc}$. We selected candidate pre-main-sequence stars using isochrones in the $M_\mathrm{G}$ vs. \gbp$-$\grp\  colour--magnitude diagram. We used HDBSCAN to find groups in the selected sample, and found 13 stellar groups, consisting of 874 stars. These 13 groups, located between 830--940\,pc, are subsystems of Cepheus~OB2. 355 of the clustered stars are classified as candidate YSOs by \citet{Marton2022}. Based on \wise\ data we identified one Class~I source, sixty-two Class~II sources and one with transitional disc, according to the classification of \citet{Koenig2014}.

We found that star formation in the volume of the association started some 10 million years ago. In addition to the evolved high-mass stars, the oldest subsystem Cep~OB2a contains four groups of low- and intermediate-mass stars. The tangential velocities suggest the expansion of the group containing NGC\,7160, however the expansion of the whole Cep~OB2a cannot be proved.

In addition to the historical younger subsystem Cep~OB2b, Trumpler~37, several other, 2--8\,Myr old groups can be found in the volume of the association. Most of them might have been formed under the influence of the expanding bubble. Spatial and kinematic structure of the region suggest various means of interaction between the star-forming clouds and the expanding bubble.  
The tangential velocities show that the parent clouds of the young clusters Trumpler~37, Alessi--Teutsch~5 and [BDS2003]~30 (Groups~11, 12 and 13) have been accelerated by the expanding ionization front of Cep~OB2a. Other groups reveal places where the expanding ionization front encountered  ambient clouds (Groups~1, 8, and 10, aka NGC\,7129, S140, and UPK\,169). The age of the cluster Pismis--Moreno~1 (Group 8, S140) suggests that formation of this group was probably independent of the expanding bubble. The relation of S129 to the Cepheus Bubble is uncertain, because of the uncertain distance of the exciting star and the discrepancy of the age of the central cluster and the exciting star. 

\section*{Acknowledgements}
We thank the anonymous referee for the careful and insightful review of our manuscript. We thank Lajos Balázs for his help in statistics. This work has made use of data from the European Space Agency (ESA) mission {\it Gaia} (\url{https://www.cosmos.esa.int/gaia}), processed by the {\it Gaia} Data Processing and Analysis Consortium (DPAC,
\url{https://www.cosmos.esa.int/web/gaia/dpac/consortium}). Funding for the DPAC has been provided by national institutions, in particular the institutions participating in the {\it Gaia} Multilateral Agreement. This work was supported by the ESA PRODEX Contract nr. 4000129910. For this work we have used Astropy \citep{astropyI, astropyII}, Matplotlib \citep{matplotlib}, Pandas \citep{pandas}, scikit-learn \citep{scikit-learn}, SciPy \citep{scipy}, TOPCAT \citep{topcat}.

\section*{Data availability}
The data of Table~\ref{tab:hms}, \ref{tab:groups} and \ref{tab:groupmembers} are available in the online supplementary material and at the CDS.

\bibliographystyle{mnras}
\bibliography{cepheusbubble}
\bsp	

\appendix
\section{Figures of stellar groups}
In this section, we show the spatial, distance, tangential velocity distribution, colour--magnitude and colour--colour diagrams of Groups~2--13. For a detailed description, see Fig.~\ref{fig:group1}. 
\begin{landscape}
\begin{figure}
    \centering
    \begin{minipage}[t]{0.47\linewidth}
        \centering
        \includegraphics[width=\textwidth]{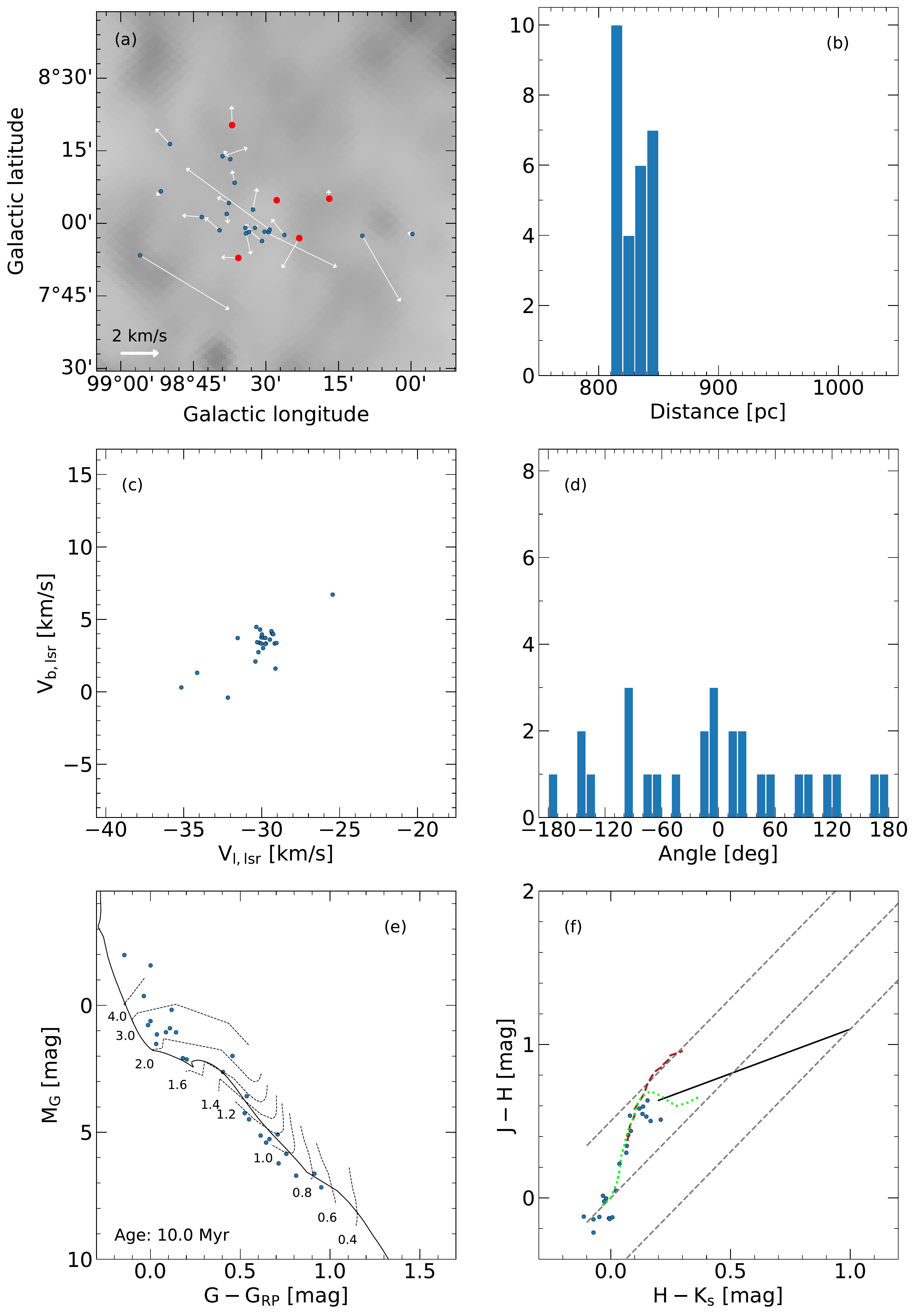}
        \caption{Same as Fig.~\ref{fig:group1} for Group~2.}
        \label{fig:group2}
    \end{minipage}
    \hfill
    \begin{minipage}[t]{0.47\linewidth}
        \centering
        \includegraphics[width=\textwidth]{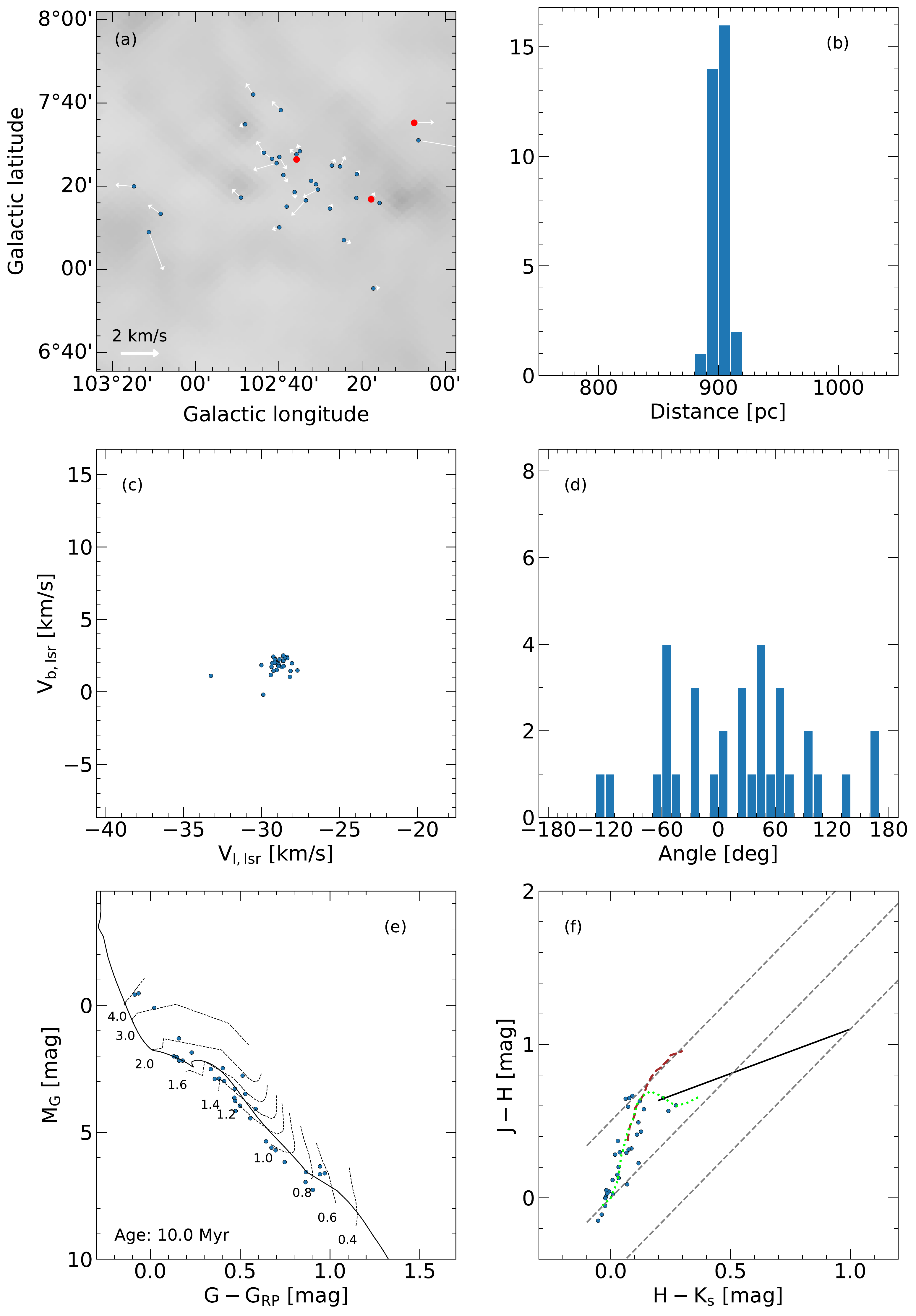}
        \caption{Same as Fig.~\ref{fig:group1} for Group~3.}
        \label{fig:group3}
    \end{minipage}%
\end{figure}
\clearpage
\newpage
\begin{figure}
    \centering
    \begin{minipage}[t]{0.47\linewidth}
        \centering
        \includegraphics[width=\textwidth]{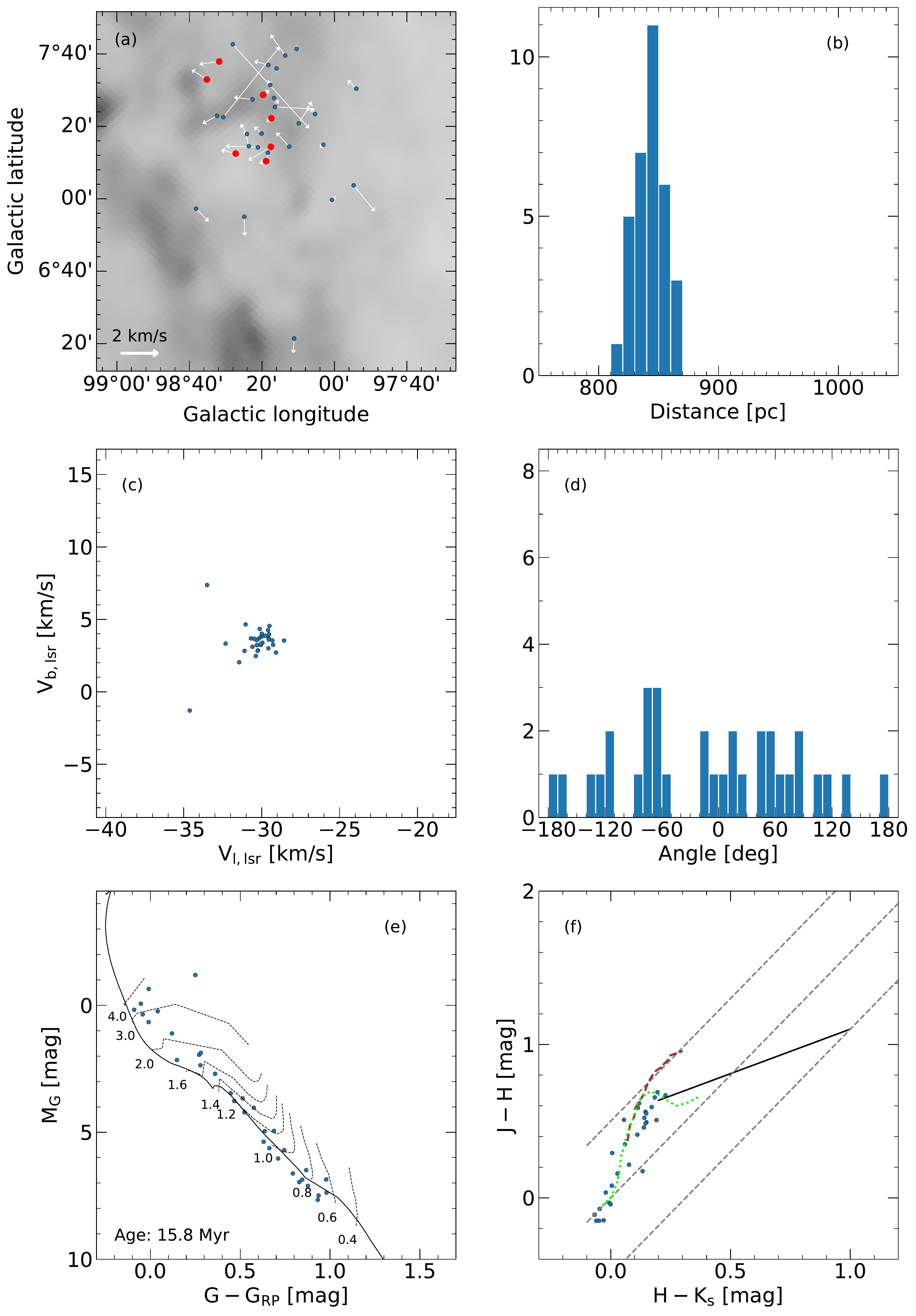}
        \caption{Same as Fig.~\ref{fig:group1} for Group~4.}
        \label{fig:group4}
    \end{minipage}
    \hfill
    \begin{minipage}[t]{0.47\linewidth}
        \centering
        \includegraphics[width=\textwidth]{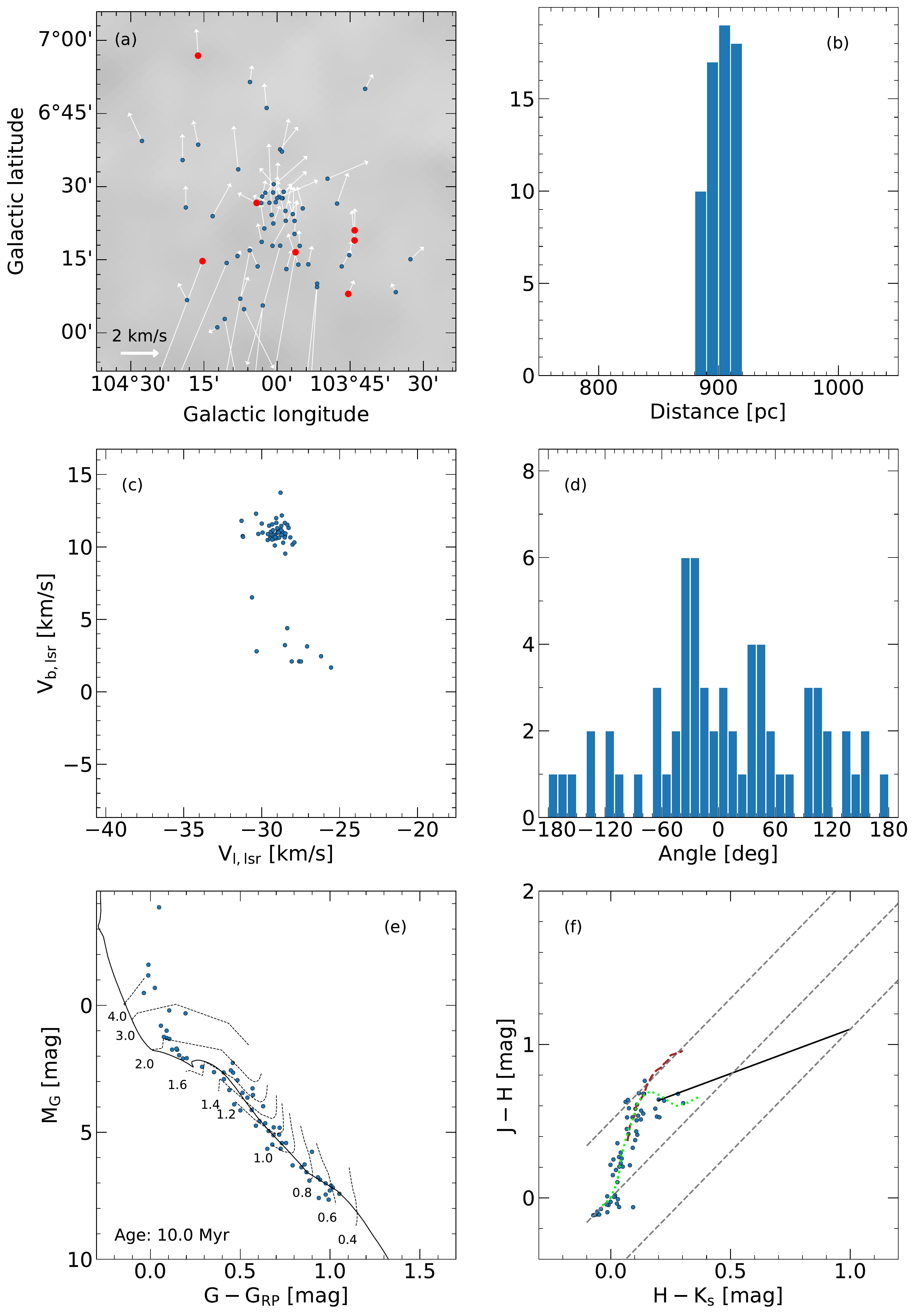}
        \caption{Same as Fig.~\ref{fig:group1} for Group~5.}
        \label{fig:group5}
    \end{minipage}%
\end{figure}
\clearpage
\newpage
\begin{figure}
    \centering
    \begin{minipage}[t]{0.47\linewidth}
        \centering
        \includegraphics[width=\textwidth]{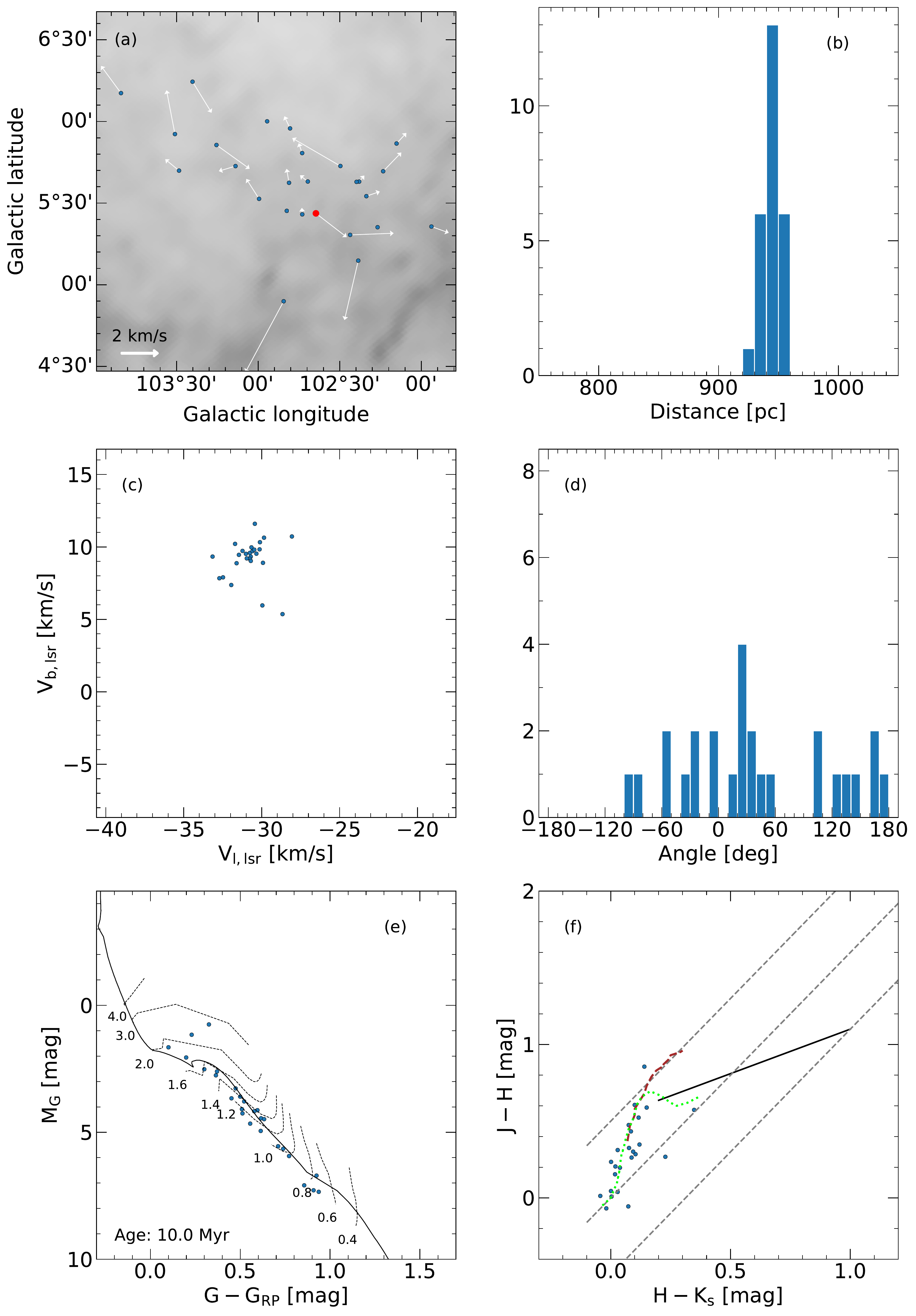}
        \caption{Same as Fig.~\ref{fig:group1} for Group~6.}
        \label{fig:group6}
    \end{minipage}
    \hfill
    \begin{minipage}[t]{0.47\linewidth}
        \centering
        \includegraphics[width=\textwidth]{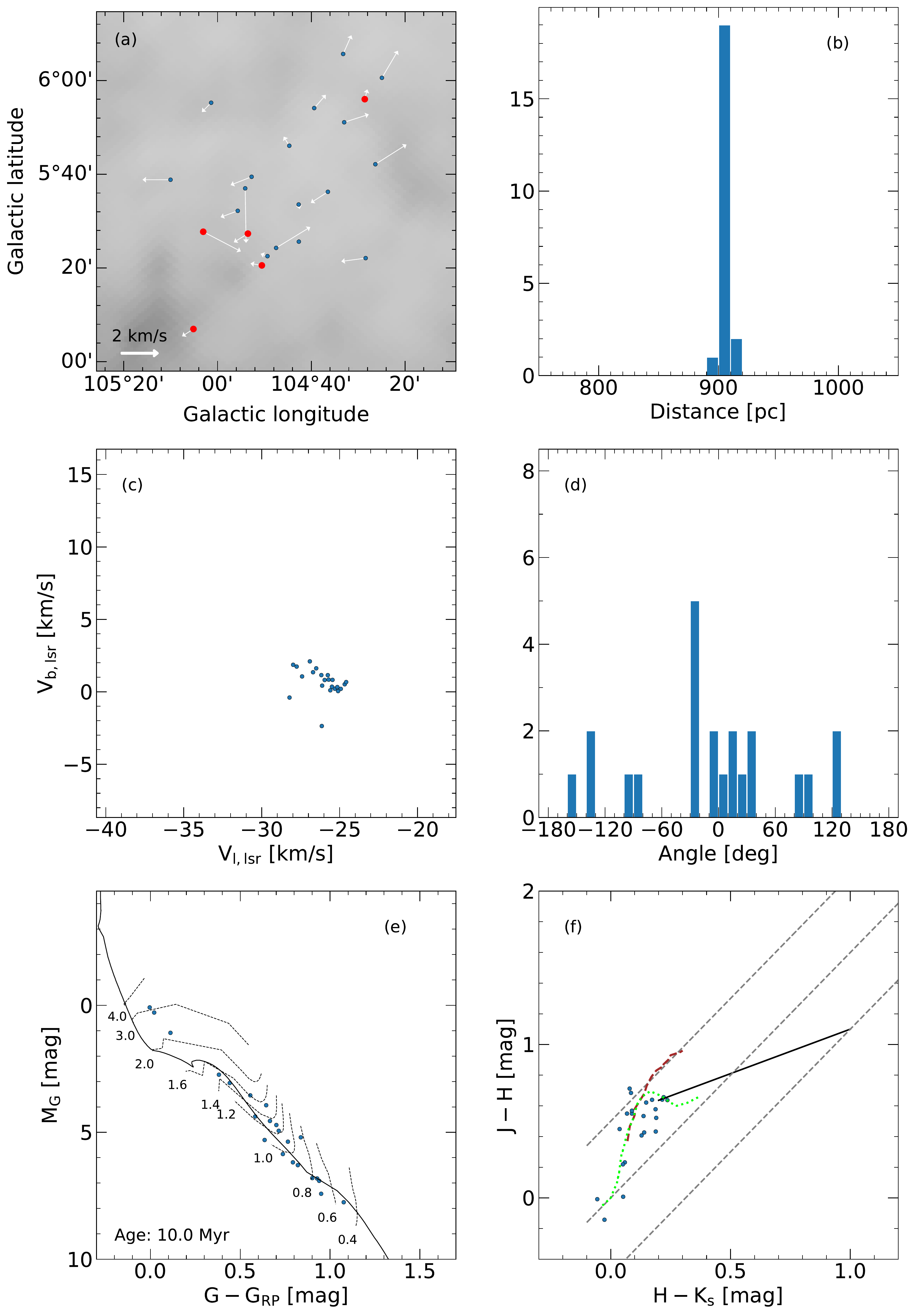}
        \caption{Same as Fig.~\ref{fig:group1} for Group~7.}
        \label{fig:group7}
    \end{minipage}%
\end{figure}
\clearpage
\newpage
\begin{figure}
    \centering
    \begin{minipage}[t]{0.47\linewidth}
        \centering
        \includegraphics[width=\textwidth]{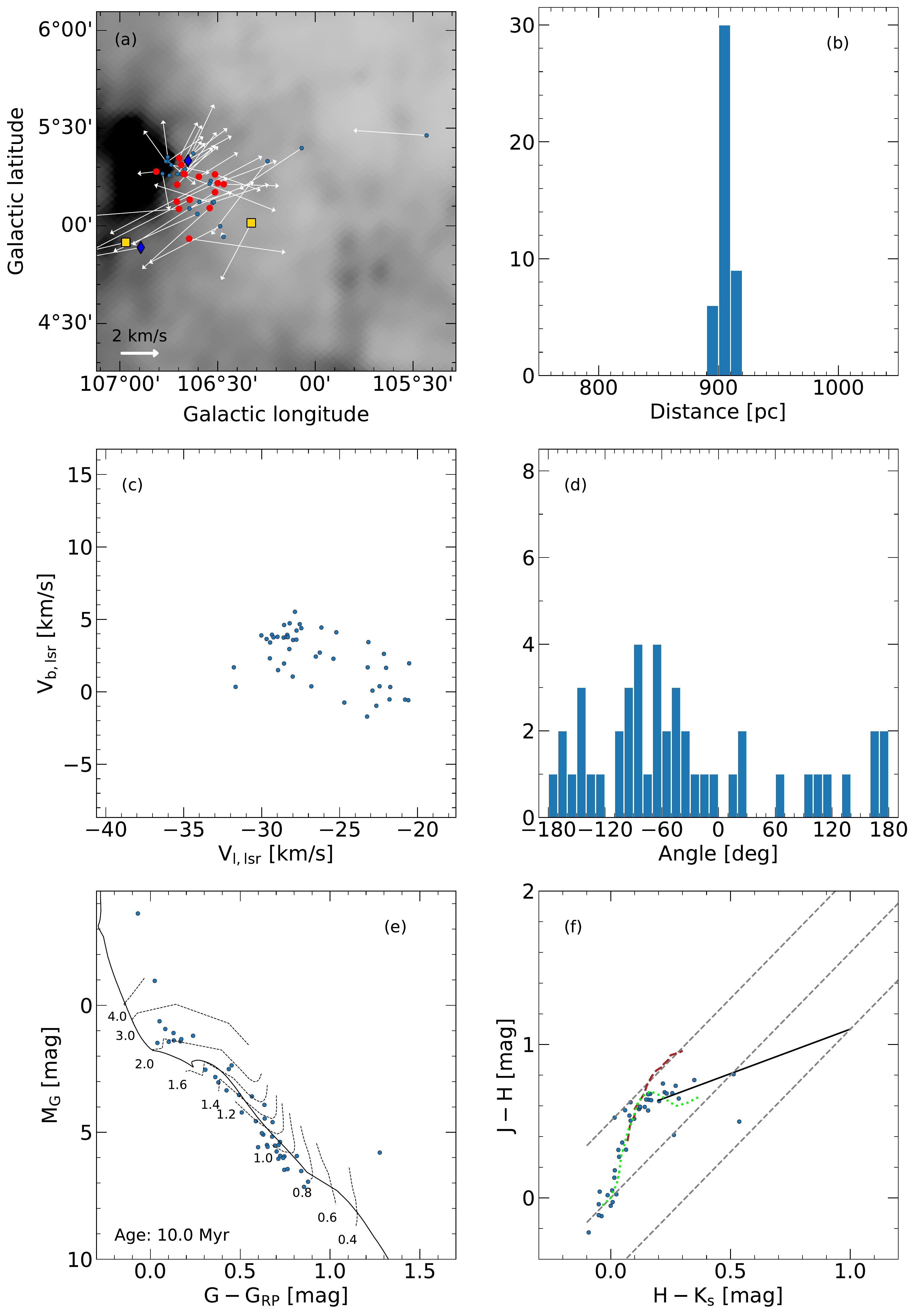}
        \caption{Same as Fig.~\ref{fig:group1} for Group~8. The blue diamonds show the Class II sources identified with \tmass\ data.}
        \label{fig:group8}
    \end{minipage}%
    \hfill
    \begin{minipage}[t]{0.47\linewidth}
        \centering
        \includegraphics[width=\textwidth]{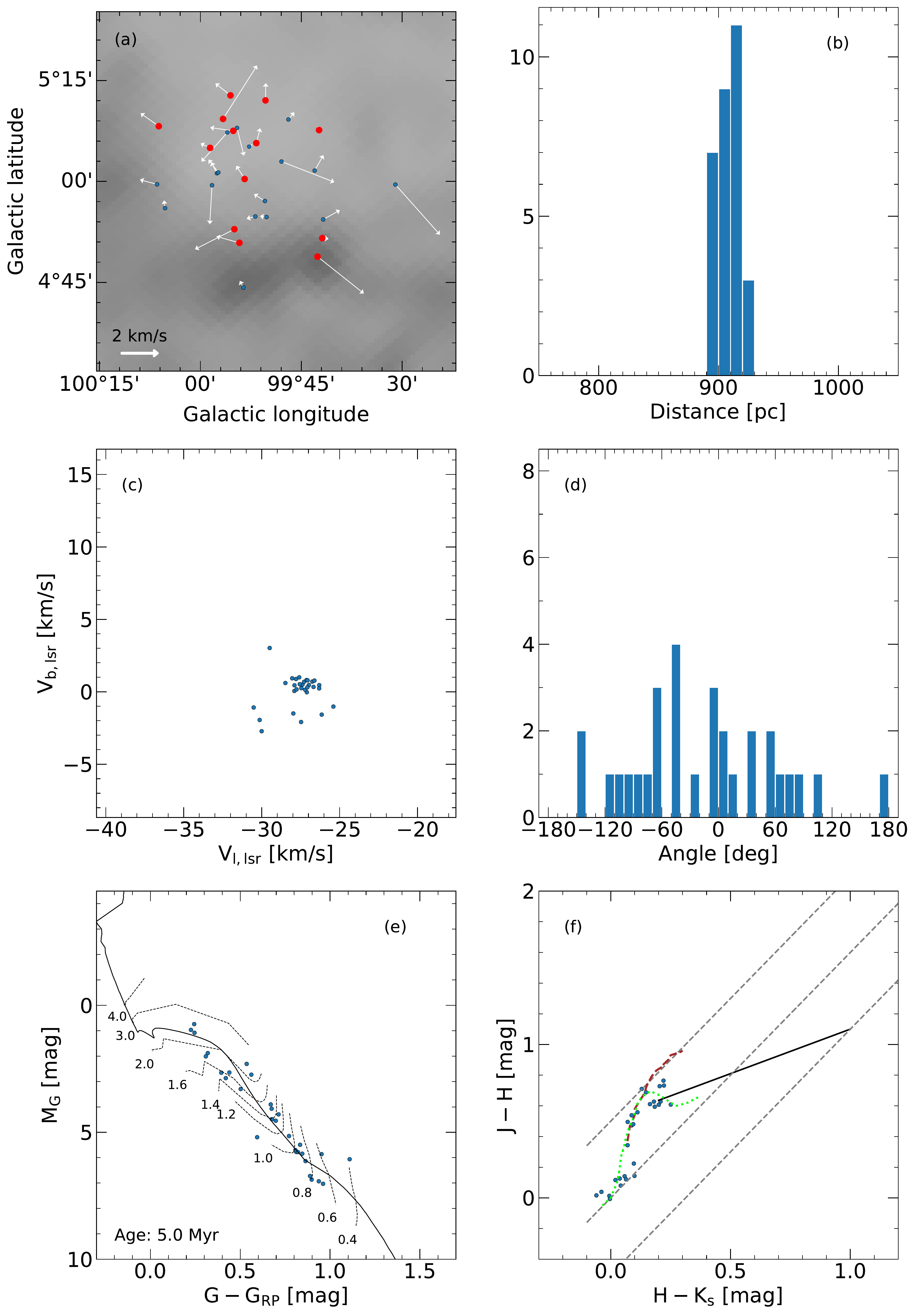}
        \caption{Same as Fig.~\ref{fig:group1} for Group~9.}
        \label{fig:group9}
    \end{minipage}
\end{figure}
\clearpage
\newpage
\begin{figure}
    \centering
    \begin{minipage}[t]{0.47\linewidth}
        \centering
        \includegraphics[width=\textwidth]{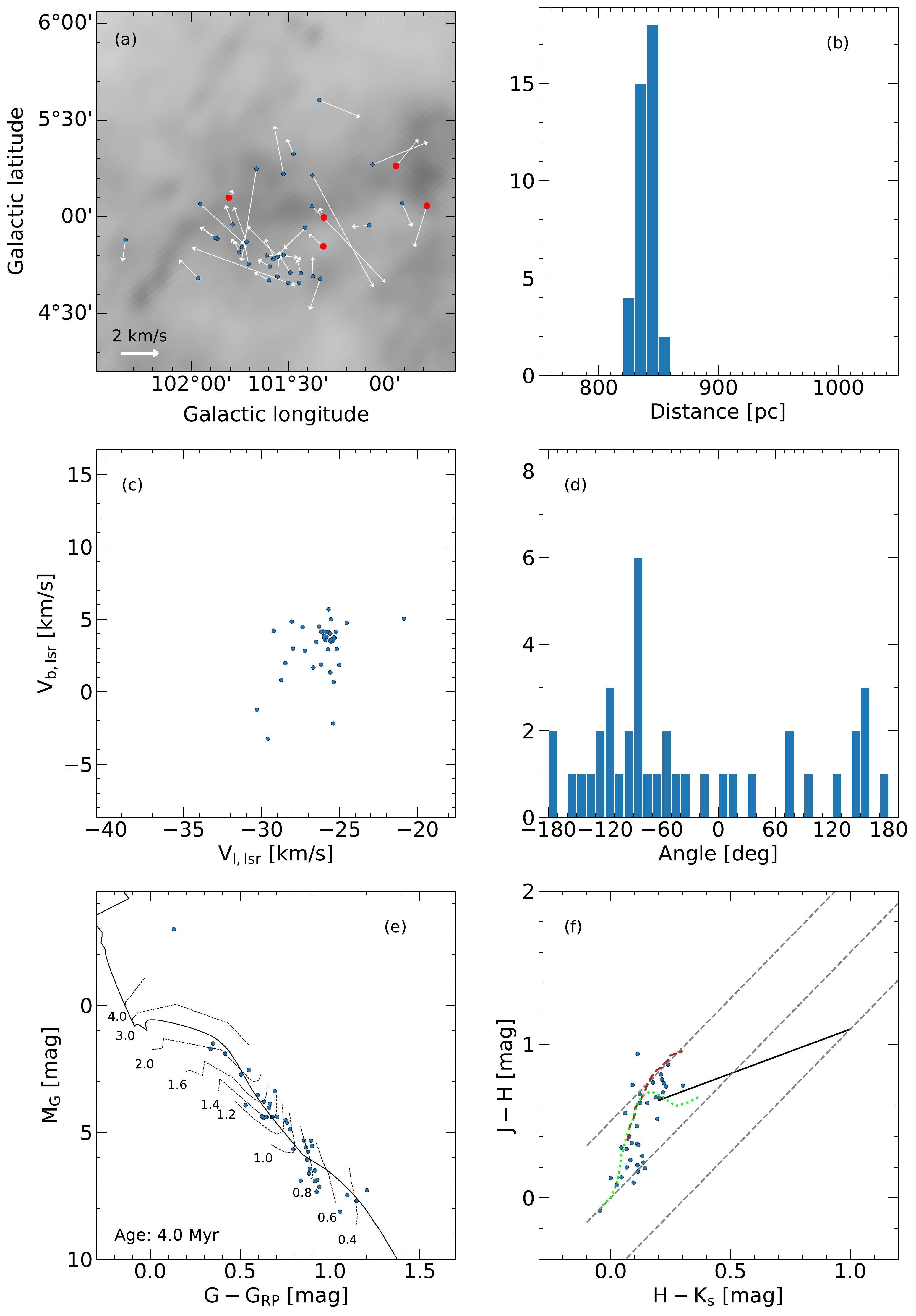}
        \caption{Same as Fig.~\ref{fig:group1} for Group~10.}
        \label{fig:group10}
    \end{minipage}%
    \hfill
    \begin{minipage}[t]{0.47\linewidth}
        \centering
        \includegraphics[width=\textwidth]{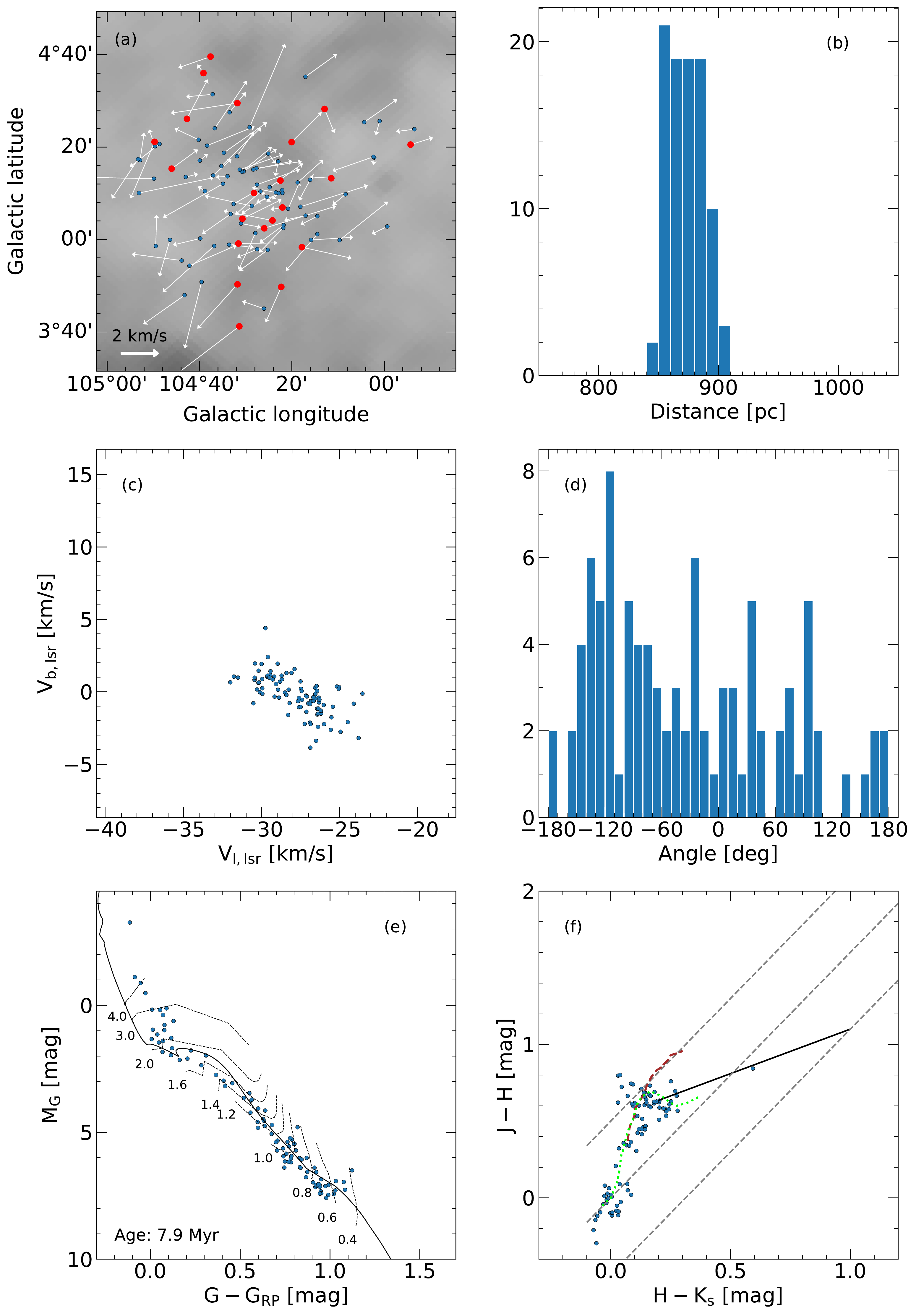}
        \caption{Same as Fig.~\ref{fig:group1} for Group~11.}
        \label{fig:group11}
    \end{minipage}
\end{figure}
\clearpage
\newpage
\begin{figure}
    \centering
    \begin{minipage}[t]{0.47\linewidth}
        \centering
        \includegraphics[width=\textwidth]{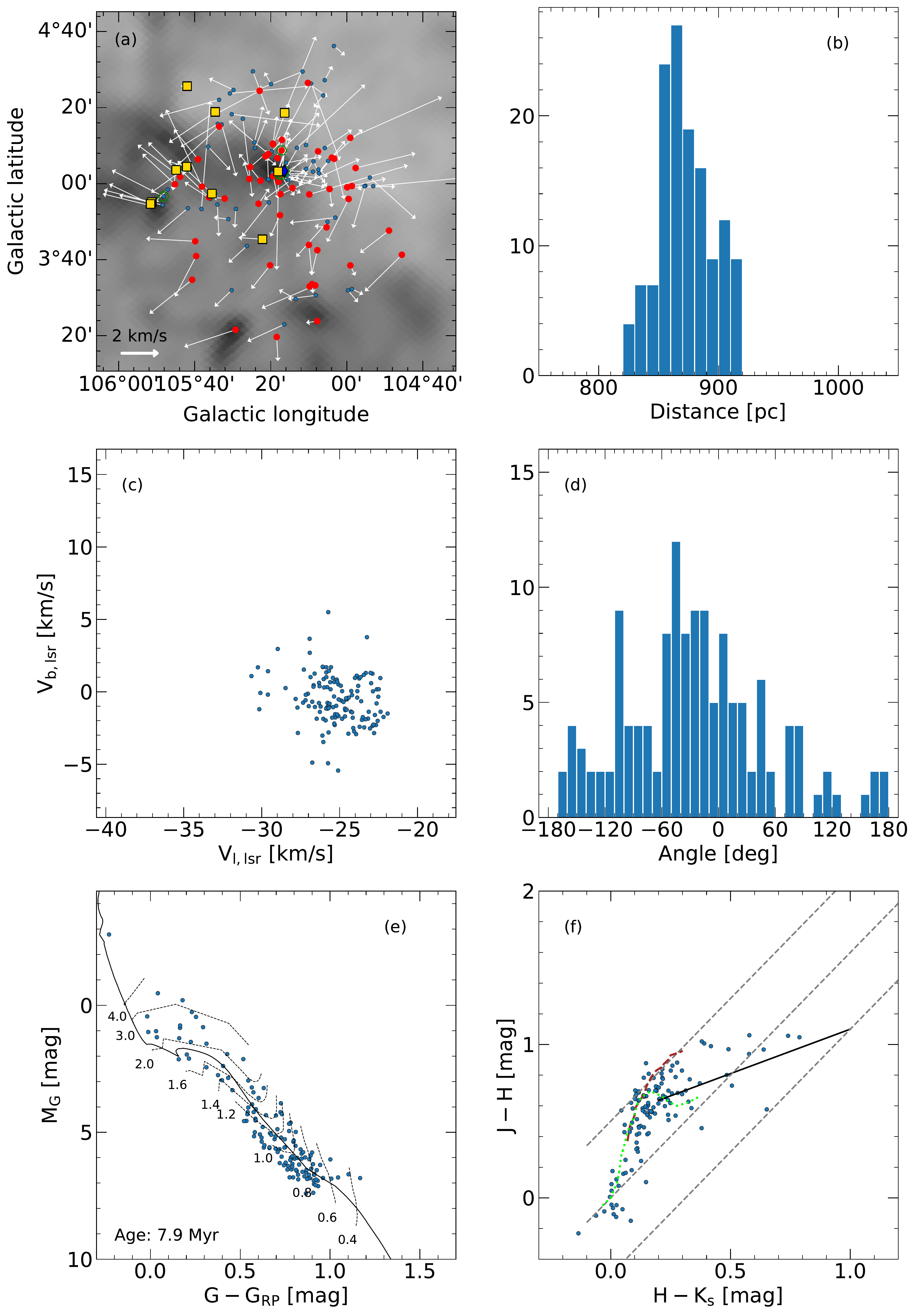}
        \caption{Same as Fig.~\ref{fig:group1} for Group~12. YSOs from \citet{Szegedi2019} are overplotted with green hexagons.}
        \label{fig:group12}
    \end{minipage}%
    \hfill
    \begin{minipage}[t]{0.47\linewidth}
        \centering
        \includegraphics[width=\textwidth]{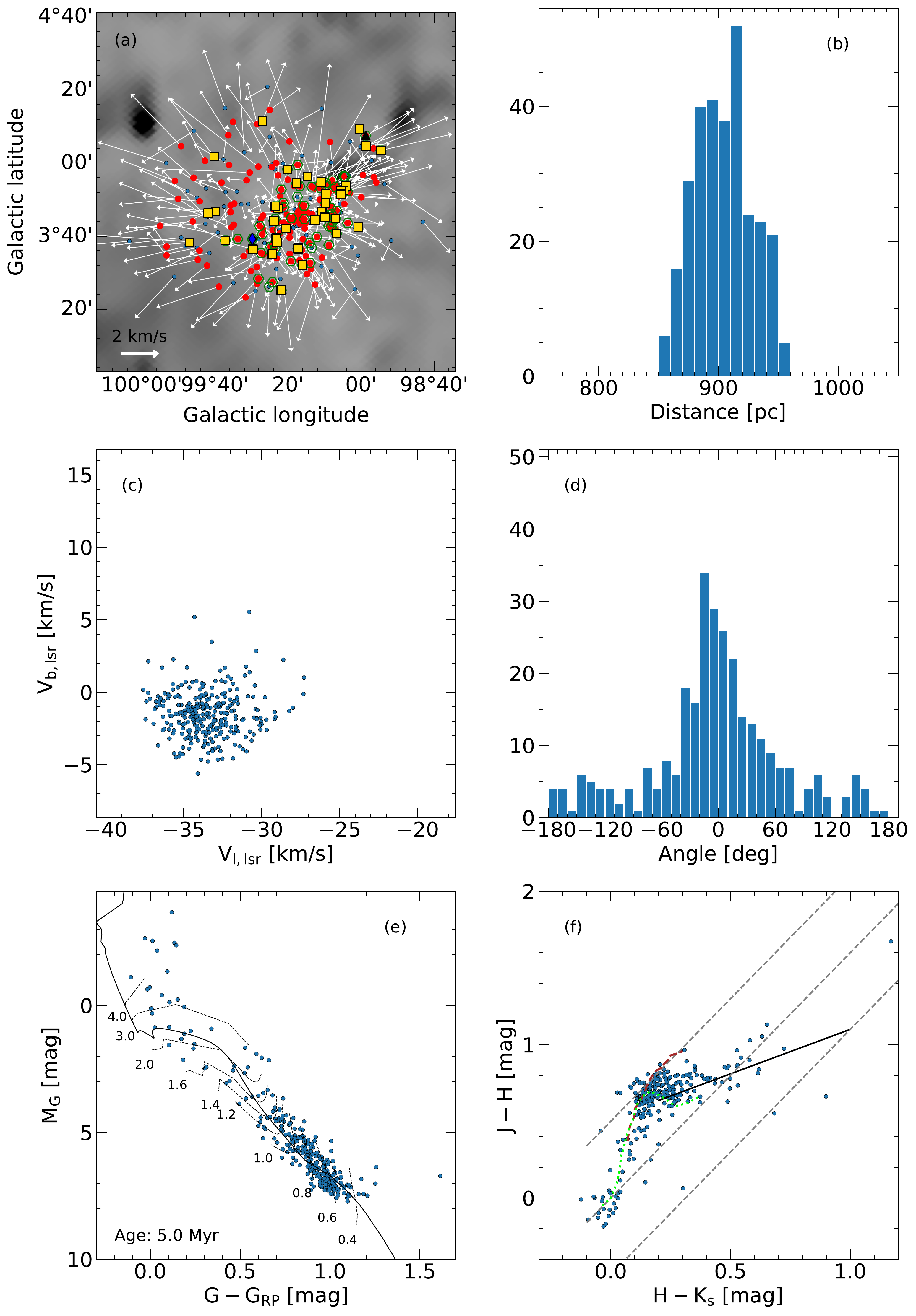}
        \caption{Same as Fig.~\ref{fig:group1} for Group~13. The black triangle indicate the transitional disc-bearing source identified by the \wise\ colour indices. YSOs from \citet{Dias2002} and \citet{SA2005} are overplotted with green hexagons.}
        \label{fig:group13}
    \end{minipage}
\end{figure}

\end{landscape}

\end{document}